\documentclass{JHEP3}
\usepackage{graphicx}
\usepackage{epsfig}
\usepackage{amssymb}
\usepackage{amsmath}
\usepackage{float}

\newcommand{\bea}{\begin{eqnarray}}
\newcommand{\eea}{\end{eqnarray}}
\newcommand{\bean}{\begin{eqnarray*}}
\newcommand{\eean}{\end{eqnarray*}}
\newcommand{\nn}{\nonumber \\}

\def\O #1{\overline{#1}}

\def\W #1{\widetilde{#1}}
\def\WH #1{\widehat{#1}}

\def\braket#1{\left\langle #1 \right\rangle}

\def\ket#1{\left| #1\right\rangle}
\def\gb #1{ \left\langle #1 \right]}
\def\tgb #1{ \left[ #1 \right\rangle}

\def\wh{\widehat}

\def\a{{\alpha}}

\def\la{\lambda}

\def\vev{\braket}
\def\tgb #1{ \left[ #1 \right\rangle}
\def\bket#1{\left| #1\right]}
\def\bvev#1{\left[ #1 \right]}
\def\Spaa{\vev}
\def\Spbb{\bvev}
\def\Spab{\gb}
\def\Spba{\tgb}

\def\Label#1{\label{#1}%
  \smash{\hbox to0pt{\raise1ex\hbox{\tiny[#1]}\hss}}}

\title{Roots of Amplitudes}

\author{Bo Feng${}^{a,b,c}$, Yin Jia ${}^{a}$, Hui Luo ${}^{a}$, Mingxing Luo ${}^{a}$,  \\
$^a$\small Zhejiang Institute of Modern Physics, Zhejiang
University, Hangzhou, 310027, P. R. China\\$^b$\small Center of
Mathematical Science, Zhejiang University, Hangzhou, China \\
$^c$\small Kavli Institute for Theoretical Physics China, CAS,
Beijing 100190, China}

\date{\today}
\abstract{In a recent paper \cite{Benincasa:2011kn}, boundary
contributions in BCFW recursion relations have been related to roots
of amplitudes. In this paper, we make several analyses regarding to
this problem. Firstly, we use different ways to re-derive boundary
BCFW recursion relations given in \cite{Benincasa:2011kn}. Secondly,
we generalize factorization limits to $z$-dependent ones, where
information of roots is more transparent. Then, we demonstrate our
analysis with several examples. In general, relations from
factorization limits cannot guarantee to find explicit expressions
for roots. }


\begin{document}

\newpage

\section{Introduction}

The standard method to calculate scattering amplitudes in quantum
field theory relies on Feynman diagrams. However, such computations
become extremely complex with increasing external particles.
Naturally, more efficient methods are desired. Among all developed
methods, the on-shell method is particularly promising.
Unitarity-cut method
\cite{Landau:1959fi,Bern:1994cg,Bern:1994zx,Britto:2004nc,Britto:2005ha,Anastasiou:2006jv,
Forde:2007mi, Bern:2010tq} is very powerful for the one-loop as well
as higher loop calculations. On-shell recursion relations
\cite{Britto:2004ap, Britto:2005fq} are not only very useful for
practical calculations but also helpful to understand many properties
of quantum field theories.

In the derivation of on-shell recursion relation, one expresses the
amplitude as an analytic function $M(z)$ of single complex variable
$z$ with momenta deformation  $p_i\to p_i-zq$, $p_j\to p_j+z q$,
where  $q^2=q\cdot p_i=q\cdot p_j=0$. The function $M(z)$ has single
poles at finite positions of $z$ as well as possibly a multiple pole
at $z=\infty$ ( i.e., the boundary). The behavior of $M(z)$ around
finite single poles can be analyzed  by factorization properties.
The behavior of $M(z)$ around $z=\infty$  is not well understood. In
many examples, with proper choice of deformed pair $(p_i, p_j)$,
$M(z)\to 0$ when $z\to \infty$, thus boundary contributions can be
avoided (i.e., $z=\infty$ is not a pole). However, if $M(z)\to C_0+
C_1 z+...C_k z^k$ with $k\geq 0$, $z=\infty$ is a pole. To get
amplitudes under these circumstances, we have to know the value of
$C_0$ which yields the  boundary contribution.

There are field theories in which nontrivial boundary contributions cannot be avoided, no matter which deformed pair is chosen.
Familiar examples are the $\la \phi^4$ theory and theories with Yukawa couplings.
Several proposals have been made to deal with boundary contributions.
The first \cite{Paolo:2007,Boels:2010mj} is to add auxiliary fields
such that boundary contributions for the enlarged theory are zero.
By proper reduction one gets desired amplitudes.
The second \cite{Feng:2009ei,Feng:2010ku,Feng:2011tw} is
to carefully analyze Feynman diagrams and isolate boundary contributions within them.
With these information, boundary contributions can be evaluated directly or recursively.
The third \cite{Benincasa:2011kn} is to express boundary contributions in terms of roots of amplitudes.
Generically, we can write
\bea M(z)=\sum_{\a=1}^{N_p} {a_\a\over z-z_\a} +\sum_{l=0}^v C_l z^l
~~~\label{Az-form-gen}\eea
where $a_\a$ can be calculated by using factorization properties while
the $C_l$'s are related to boundary behaviors.
$M(z)$ can be rewritten as $P_{N_p+v}(z)/ \prod_{\a} (z-z_\a)$.
 $P_{N_p+v}(z)$ is a polynomial of $z$ of degree $N_p+v$, so there are $N_p+v$
roots of $M(z)$.
Expressing $C_l$'s in terms of $v+1$ roots, one relates boundary contributions to the latter.

This translation to roots is nice.
Following  upon that, Benincasa and Conde \cite{Benincasa:2011pg} have discussed the
extension of the constructible notion initiated in \cite{Paolo:2007}.
In this paper, we would like to explore several issues for this proposal.
The first issue is how practical this procedure is for real calculations.
The second one is as the following.
Starting from a physical amplitude $M(z=0)$, after the $z$-deformation
we will arrive at the form (\ref{Az-form-gen}) with given power $v$ and unique $C_l$.
However, one can always add an arbitrary polynomial $zf(z)$ to get a new function $\W M(z)=M(z)+z f(z)$.
Both $\W M(z)$ and $M(z)$ yield the same amplitude when $z=0$,
but $\W M(z)$ may have a different set of roots from $M(z)$.
This ambiguity matters especially when we try to construct amplitudes recursively,
i.e., starting from lower-point amplitudes to find higher-point amplitudes.
Certain principle will be needed to infer roots of $n$-point amplitudes
if we know only roots of $m$-point amplitudes with $m<n$.

In section 2, we re-derive the on-shell recursion relation with boundaries  in \cite{Benincasa:2011kn} via a new method.
The key here is the shuffling of roots.
Presented in section 3 are the factorization limits that can be used to deal with roots.
Following this brief discussion, we analyze the $z$-parameterized factorization limit carefully in section 4.
We will consider  poles with and without $z$-dependence, respectively
and then use different efficient limits to construct consistent conditions from factorization and boundary BCFW  relations.
One obtains some information about roots under these limits, but not enough to determine them precisely in general.
Our analysis is thus inconclusive.
In section 5, several examples are calculated to demonstrate these general discussions.
Finally, we conclude in section 6.

\section{A new derivation of on-shell recursion relations with boundary contributions}

We present in this section a new derivation of on-shell recursion relations
with boundary contributions, in contrast with the one given in \cite{Benincasa:2011kn}.
The BCFW recursion relation with boundary contributions can be written as
\bea M_n(z)=\sum_{k\in \mathcal{P}^{(i,j)}}{M_L(z_k)M_R(z_k)\over
P_k^2(z)}+C_0+\sum_{l=1}^{v}C_l z^l~,~~~\label{M-BCFW-z1}\eea
where we have assumed that $i\in k$ so $P_k^2(z)= (-2P_k\cdot q)
(z-z_k)$ with $z_k=P^2_k / 2P_k \cdot q$.
Pulling all denominators together, one has
\bea M_n(z)= c{\prod_{l}(z-w_l)^{m_l}\over
\prod_{k=1}^{N_p}P_k^2(z)}~,~~~\sum_l
m_l=N_z=N_p+v~,~~~\label{bcfw-factorization}\eea
here $w_l$ are roots of $M_n(z)$.

Unlike results without boundary contributions,
(\ref{M-BCFW-z1}) has single poles at finite locations of $z$ and a pole at  $z=\infty$ of degree $v+1$ as well.
To completely determine $M_n(z)$, we need to determine not
only residues of single poles at finite locations, but also
coefficients related to the pole of degree $v$ at $z=\infty$.

With $N_z\geq N_p$  in (\ref{bcfw-factorization}), we can split roots into two groups ${\cal I},{\cal J}$
 with number of roots $n_{\cal I}$ and $n_{\cal J}$ ($N_z=n_{\cal I}+n_{\cal J}$) respectively.
 If $n_{\cal I}< N_p$, we can write ($w_l$ is a root of multiplicity $m_l$)
\bea c{\prod_{l=1}^{n_{\cal I}}(z-w_l)\over
\prod_{k=1}^{N_p}P_k^2(z)}=\sum_{k=1}^{N_p} {c_k\over
P_k^2(z)}~,~~~\label{Np-zero-pole}\eea
where $c_k$'s are unknown $z$-independent coefficients.
Plugging (\ref{Np-zero-pole}) back to (\ref{bcfw-factorization}), we have
\bea M_n(z)= \sum_{k=1}^{N_p} {c_k\over
P_k^2(z)}\prod_{l=1}^{n_{\cal J}}(z-w_l)~.~~~~\label{M-zero-z1}\eea
Performing a contour integration of (\ref{M-BCFW-z1}) and (\ref{M-zero-z1}) around a single pole $z_k$,
one obtains
\bea  {M_L(z_k)M_R(z_k)\over (-2P_k\cdot q)}= {c_k\over (-2P_k\cdot
q)}\prod_{l=1}^{n_{\cal J}}(z_k-w_l), ~~\Longrightarrow
c_k={M_L(z_k)M_R(z_k)\over \prod_{l=1}^{n_{\cal J}
}(z_k-w_l)}.~~~\label{ck-sol}\eea
Plugging (\ref{ck-sol}) into (\ref{M-zero-z1}) , we have
\bea M_n(z)= \sum_{k\in \mathcal{P}^{(i,j)}} {M_L(z_k)M_R(z_k)\over
P_k^2(z)}\prod_{l=1}^{n_{\cal J}}{(z-w_l)\over
z_k-w_l}~~~~\label{M-zero-z2}\eea
which is similar to what given in \cite{Benincasa:2011kn}, but with new features.

In (\ref{M-zero-z2}), the splitting of roots into two groups is arbitrary as long as $n_{\cal I}< N_p$.
If  $n_{\cal I}\leq  N_p-2$, there will be extra consistent relations.
Taking $n_{\cal I}=N_p-2$ and expanding (\ref{M-zero-z2}) into the form of (\ref{M-BCFW-z1}), one has
\bea M_n(z)= \sum_{k\in \mathcal{P}^{(i,j)}} {M_L(z_k)M_R(z_k)\over
(-2P_k\cdot q)(z-z_k)}\left( {(z-z_k)^{v+2}\over \prod_l (z_k-w_l)}
+{ (z-z_k)^{v+1}\sum_l (z_k-w_l) \over \prod_l (z_k-w_l)}
+...+1\right)~.~~~~\label{M-zero-z3}\eea
The coefficient of the $z^{v+1}$ term is
$\sum_{k\in\mathcal{P}^{(i,j)}} M_L(z_k)M_R(z_k)/ (-2P_k\cdot q)\prod_l (z_k-w_l)$.
It should be zero and this results in a consistent condition.
To avoid such extra  consistent conditions and to deal with only
minimum number of roots, we will take $n_{\cal I}=N_p-1$ from now on and obtain
\bea M_n(z)= \sum_{k\in \mathcal{P}^{(i,j)}} {M_L(z_k)M_R(z_k)\over
P_k^2(z)}\prod_{l=1}^{v+1}{(z-w_l)\over
z_k-w_l}~,~~~~\label{Mz-new}\eea
which is the expression presented in \cite{Benincasa:2011kn} and will be the starting point of most our discussion.

Also, the $v+1$ roots $w_l$ in (\ref{Mz-new}) can be chosen arbitrarily from the total $N_z$ roots.
For practical purposes, $w_l$ should be chosen with certain discretion, instead of being left totally arbitrary.
This is related to the issue raised in the introduction, namely, the arbitrariness in defining $M_n(z)$.

Having established (\ref{Mz-new}),  we can get the boundary BCFW recursion relation by setting $z=0$
\bea M_n= \sum_{k\in \mathcal{P}^{(i,j)}} {M_L(z_k)M_R(z_k)\over
P_k^2}\prod_{l=1}^{v+1}{w_l\over w_l-z_k}~.~~~~\label{Mz=0-new}\eea
The coefficients $C_l$ in (\ref{M-BCFW-z1}) can be  read out by expanding (\ref{Mz-new}).
Notice that $(z-w_l)/(z_k-w_l)= 1+ (z-z_k)/(z_k-w_l)$ and
\bea \prod_{l=1}^{v+1}\left({z-z_k\over z_k-w_l}+1\right) =1+
\sum_{s=1}^{v+1} (z-z_k)^ s{\sum}^{'} {1\over
\prod_{\sigma=1}^s (z_k-w_\sigma)}
\eea
where the sum ${\sum}^{'}$  is over all  $C_{v+1}^s$ possible selections of  $s$
$(z_k-w_\sigma)$-factors from all $(v+1)$ factors. Thus we have
\bea
C_l= \sum_{k\in \mathcal{P}^{(i,j)}} {M_L(z_k)M_R(z_k)\over (-2P_k\cdot q)}
\sum_{s=l+1}^{v+1}  \left. {d^{l}(z-z_k)^{s-1}\over dz^l}\right|_{z\to 0}
{\sum}^{'} {1\over \prod_{\sigma=1}^s (z_k-w_\sigma)},~~~l=0,1,...,v
\eea

We  now address two more points before ending this section.
Firstly, the divergent degree $v$ is a function of $n$ in general quantum field theory
(except gauge theory, gravity theory or other well-defined renormalizable theories).
If one adds an interaction vertex with arbitrary number of external fields,
the divergent degree $v$ will be modified.
Secondly, both poles and roots are important to determine tree level amplitudes.
Poles are local property and easier to determine while roots are (quasi)global property and harder to deal with.
In general, roots depend on the choice of deformed pair, helicity configuration and other detail information.

\section{Strategy to find roots}

To find the $n$-point amplitude via (\ref{Mz=0-new}),
the crucial point  is to find its roots.
Starting with the known roots of $m$-point amplitudes,
we hope to find the roots of $n$-point amplitude ($m<n$), recursively.
How to do so?

It was suggested in \cite{Benincasa:2011kn} to consider consistent conditions,
obtained from various collinear or multiple particle factorization channels.
Here the higher-point amplitude consists of the product of two lower-point amplitudes,
from which we may infer information of roots under factorization limits.

In \cite{Benincasa:2011kn},
the factorization limit is always taken for physical amplitude $M_n(z=0)$ given in (\ref{Mz=0-new}).
Since the deformation is on-shell, i.e., $M_n(z)$ is an on-shell amplitude for every $z$,
the factorization limit can actually be taken for $z$-parameterized amplitude $M_n$ given in (\ref{Mz-new}).
In other words, we should have the following consistent condition  for any $z$
\bea \lim_{P_\a^2(z)\to 0} P_\a(z)^2 \sum_{k\in
\mathcal{P}^{(i,j)}}{M_L(z_k)M_R(z_k)\over
P_k^2(z)}\prod_{l=1}^{v+1}{(z-w_l)\over (z_k-w_l)} = M_L(z)
M_R(z)~.~~~~\label{Gen-cons}\eea
Conditions (\ref{Gen-cons}) are much stronger, because both sides
are functions of $z$, not merely their values at $z=0$, to be
compared. However, condition $P_\a(z)^2=0$ holds only for a specific
value of $z$ in general, as $P_{\a}^2(z)=P_\a^2-2z P_\a\cdot q$ if
$p_i$ is to not allow to change with $z$. Thus conditions
(\ref{Gen-cons}) can not be imposed for general channels, except for
two particle channels and channels do not contain $p_i, p_j$. These
issues are to be investigated carefully in the next section.

The need of careful distinction between two factorization limits in (\ref{Gen-cons}) relates closely to the second issue raised in the introduction.
In many examples, due to constraints from $z$-parameterized factorization limits,
the $z$-dependent amplitude has not much freedom to add a polynomial $z f(z)$.
At the same time, conditions (\ref{Gen-cons}) constrain the roots $w_t$ selected in (\ref{Mz-new}) and (\ref{Mz=0-new}).

Constraints from factorization limits provide some information of roots,
but these constraints are not enough to get complete answers for roots, as to be shown in the
example of the six-gluon amplitude.
Our analysis is thus inconclusive.

\section{$z$-parameterized factorization limits}

We now discuss $z$-parameterized factorization limits
(\ref{Gen-cons}) carefully, following
\cite{Schuster:2008nh,Benincasa:2011kn}. We will mainly concern the
collinear limit ( i.e., two particle channel), although we know that
for some theories,  there is no two particle channel (i.e., there is
no on-shell three-point amplitude) with BCFW-recursion relation,
such as in the $\la \phi^4$ theory.

We will start by fixing notations and collecting some useful results for latter discussion,
then proceed by the analysis of various channels one by one.

\subsection{Conventions  and useful results}

Choosing the $(i,j)$-pair the  deformation is given as
\bea p_i\to p_i-z q,~~~~p_j\to p_j+z q,~~~~q^2=q\cdot p_i=q\cdot
p_j=0.~~~\label{BCFW-q-defor} \eea
This deformation (\ref{BCFW-q-defor}) works for massive or massless theories.
Here we consider only massless theories, thus $q$ can be solved directly.
We will choose $q=\la_i \W \la_j$ by the $\Spba{i|j}$-deformation
\bea \W \la_i\to \W\la_i-z\W\la_j,~~~~~\la_j\to
\la_j+z\la_i~.~~~~\label{Spba-ij} \eea
With this convention, if particle $i$ is in the set $\a$,
$P_{\a}^2(z)=P_{iI}^2(z)=0$ will result in $z_{iI}=P_{iI}^2/ 2P_{iI}\cdot q=P_{iI}^2/ \Spab{i|P_{iI}|j}$.
Therefore the contribution from this cut to (\ref{Mz=0-new}) is
\bea T_{iI} & =& M_L (\WH p_i, -\WH P_{iI}){1\over p_{iI}^2} M_R
(\WH P_{iI}, \WH p_j) \prod_{l=1}^{v+1} {w_l\over
w_l-z_{iI}},~~~~\label{T-iI} \eea
and the $z$-parameterized amplitude $M_n(z)$ is
\bea M_n(z) & = & \sum_{I} M_L (\WH p_i, -\WH P_{iI}){1\over
P_{iI}^2(z)} M_R (\WH P_{iI}, \WH p_j)\prod_{l=1}^{v+1} {w_l-z\over
w_l-z_{iI}}~.~~~\label{general-bcfw-z} \eea
Since $p_i, P_{iI}$ take the shifted momenta in (\ref{general-bcfw-z}),
original  physical poles inside $M_L, M_R$ (e.g. $P_\a^2$, $\a\subset I$),
will be modified and become spurious ones
\bea P_{i\a}^2 & \to &  (P_{i\a} -z_{iI} q)^2= P_{i\a}^2- P_{iI}^2{
2P_{i\a}\cdot q\over 2P_{iI}\cdot q}=
{\Spab{i|P_{i\a}(P_{i\a}-P_{iI}) P_{iI}|j}\over
\Spab{i|P_{iI}|j}}~.~~~~\label{Spurious-Pole} \eea
Such spurious pole will show up at two and only two places,
corresponding to cuts $z_{iI}$ and $z_{i\a}$, and will cancel each
other. Generally, spurious poles are not physical. However, some
spurious poles can become physical. For example, if the set $\a$
contains only one single particle $k$ in (\ref{Spurious-Pole}), the
spurious pole will factorize as $\Spaa{i|k}\Spbb{k|(P_{i\a}-P_{iI})
P_{iI}|j}$, which contains the physical pole $\Spaa{i|k}=0$.

On-shell three-point amplitude will be important for late
discussions when we study the two-particle channel. As shown in
\cite{Paolo:2007}, massless three-point amplitudes are uniquely
determined by spin symmetry and Lorentz symmetry and are given by
following forms
\bea
M_3^{h}(a,b,c) & = & \Spaa{a|b}^{h_c-h_a-h_b} \Spaa{b|c}^{h_a-h_b-h_c}
 \Spaa{c|a}^{h_b-h_c-h_a},~~~if~~-h_a-h_b-h_c\geq 0~,~~\label{3-h}\\
M_3^{a}(a,b,c) & =
&\Spbb{a|b}^{-h_c+h_a+h_b}\Spbb{b|c}^{-h_a+h_b+h_c}
\Spbb{c|a}^{-h_b+h_c+h_a},~~if~~h_a+h_b+h_c\geq 0~.~\label{3-a}\eea
If the total helicity $h=\sum h_i$ is positive/negative, only anti-holomorphic/holomorphic part is nonzero.
If $h$ is zero, both are allowed.
The mass dimension of expressions (\ref{3-h}) and (\ref{3-a}) is $| h|$.
To get the overall mass dimension $+1$ for three-point amplitude,
we have to add a coupling constant $\kappa$ of the dimension $1-|h|$, i.e., $dim(\kappa)=1-|h|$.

As we mentioned earlier, poles can be divided into two categories:
those with nontrivial $z$-dependence and those without. We now
discuss these two categories one by one. Among poles without
nontrivial $z$-dependence, $P_{ij}$ may or may not exist. For
example, for color-ordered gluon amplitude, if $i,j$ are not nearby
there is no pole $P_{ij}$.

\subsection{Poles with nontrivial $z$-dependence}

Poles of this category can be denoted as $P_{iI}$ with $j\not\in I$ (or $P_{jJ}$ with $i\not\in J$).
From discussions above, we find that if $I, J$ contain two or more particles,
pole $P_{iI}$ (or $P_{jJ}$) shows up only in one term of (\ref{T-iI}) and
the $z$-independence factorization limit is trivially true as $z_{iI}\to 0$.
If we do not allow external momenta to change with $z$, this limit can not be reached for all $z$.

However, there is an exception for the $z$-dependent factorization limit.
It is the two particle channel $P_{ik}$ (or $P_{jk}$).
The reason is that the collinear limit of massless theory for two particle channel can take either $\Spaa{i|k}\to 0$ or $\Spbb{i|k}\to 0$
(but only one choice for massive theory).
Following the deformation (\ref{Spba-ij}),
\bea P_{ik}^2(z)= \Spaa{i|k}(\Spbb{k|i}-z\Spbb{k|j}) \eea
which vanishes  for all $z$ if $\Spaa{i|k}\to 0$.
Thus, we have a $z$-dependent factorization limit $P_{ik}^2(z)\to 0$ with the choice $\Spaa{i|k}\to 0$
\footnote{The $z$-independent factorization limit $\Spbb{i|k}\to 0$ will be satisfied automatically,
but there is no $z$-dependent factorization limit of $\Spbb{i|k}-z\Spbb{j|k}\to 0$}.
Similarly, one has a $z$-dependent factorization limit $P_{jk}^2(z)\to 0$ from $\Spbb{j|k}\to 0$.

Now we are going to find out where poles may show up when
$\Spaa{i|k}\to 0$ or $\Spbb{j|k}\to 0$. The first possible place is
in the cut $s_{ik}$ in (\ref{Mz=0-new}). That is, in the term
$$M_3^{(h)}(\WH i, k,-\WH P_{ik}){1\over s_{ik}}M_{n-1}(..)\prod_l {w_l\over w_l-z_{ik}}$$
 where $z_{ik}=\Spbb{k|i}/ \Spbb{k|j}$.
Because $\la_i\sim \la_k\sim \la_{P_{ik}}$ in the $M_3^h$ part,
there will be a contribution $\Spaa{i|k}^{-(h_i+h_k+h_P)}$. There is
one $\Spaa{i|k}$ from the pole $s_{ik}$, but it will be cancelled by
other $\Spaa{i|k}$  factors from $M_3^{(h)}(\WH i,k,-\WH P_{ik})$ if
$h_i+h_k+h_P\leq -1$. As a result, there may not be a $\Spaa{i|k}=0$
pole. This is true in many theories, such as pure gauge or gravity
theory, so it will be assumed from now on. Similar argument shows
that the cut $s_{jk}$ in (\ref{Mz=0-new}) does not give the pole
$\Spbb{j|k}=0$ in general. Having excluded above possibility, we are
left with only one choice: spurious poles. Fortunately, as we have
mentioned after (\ref{Spurious-Pole}), these two singularities do
appear as factor in spurious poles $T_{iI}$ with $k \in I$ (notice
that we do not have the pole $\Spbb{i|k}$ shows up in these spurious
poles for consistence) or $T_{jJ}$ with $k\in J$.

For latter purposes, we now write down two factorization limits from general principles
\bea
D_{ik}(z)=\lim_{\Spaa{i|k}\to 0} P_{ik}^2(z)M_n(z)& = & M_3^a(
i(z), k,-P_{ik}^{-h_{ik}}(z)) M_{n-1}
(P_{ik}^{h_{ik}}(z),..,j(z),..)~,~~~\label{Pik-A-z-1} \\
D_{jk}(z)=\lim_{\Spbb{j|k}\to 0} P_{jk}^2(z)M_n(z)& = & M_3^h( j(z),
k,-P_{jk}^{-h_{jk}}(z)) M_{n-1}
(P_{jk}^{h_{jk}}(z),..,i(z),..)~,~~~\label{Pjk-A-z-1} \eea
which will be compared with limits from (\ref{Mz-new}). Expressions
(\ref{Pik-A-z-1}) and (\ref{Pjk-A-z-1}) can be further simplified.
For example, in (\ref{Pik-A-z-1}) one can write $\ket{k}=
\ket{i}\Spaa{k|\mu}/ \Spaa{i|\mu}$ where $\mu$ is an arbitrary
auxiliary spinor. Thus, $ P_{ik}(z)  =  \ket{i} \left(
\bket{i}-z\bket{j}+\bket{k}{\Spaa{k|\mu}/ \Spaa{i|\mu}}\right)$ and
one can get from (\ref{3-a})\footnote{$h_i+h_k-h_{ik}\geq 0$ in this
case.}
\bea M_3^a( i(z)^{h_i}, k^{h_k},-P_{ik}^{-h_{ik}}(z)) & = &
\Spbb{i-z j|k}^{h_i+h_k-h_{ik}} (-)^{-2h_{ik}}
\left({\Spaa{k|\mu}\over \Spaa{i|\mu}} \right)^{-h_k+h_i-h_{ik}}~.
~~~~\label{Pik-A3-z-1}\eea
Similarly with $\bket{k}=\bket{j}{\Spbb{k|\mu}/ \Spbb{j|\mu}}$,
$ P_{jk}(z)  =   \left( \ket{j}+z\ket{i}+\ket{k}{\Spbb{k|\mu}/ \Spbb{j|\mu}}\right)\bket{j}$,
\footnote{$h_j+h_k-h_{jk}\leq 0$ in this case.}~
one can get\footnote{Note that while calculating $M_3^{a/h}$ we have assumed a particular ordering,
which may be different from real situation.
However, when we compare the direct factorization limit with the one obtained from (\ref{general-bcfw-z}),
the ordering ambiguity will be canceled at both sides.}
\bea M_3^h (j^{h_j}(z), k^{h_k}, -P_{jk}^{-h_{jk}}(z))=(-)^{2h_{jk}}
\Spaa{j+z i|k}^{-(h_j+h_k-h_{jk})} \left( {\Spbb{k|\mu}\over
\Spbb{j|\mu}}\right)^{h_k+h_{jk}-h_j}~~~~\label{Pjk-A3-z-1}~. \eea
%

\subsubsection{The pole $\Spaa{i|k}=0$ from the cut $P_{iI}$ with $k\in I$}

The factorization limit from (\ref{general-bcfw-z}) is given by
\bea \lim_{\Spaa{i|k}\to 0} P_{ik}^2(z)M_n(z)& = &\sum_{k\in I}
\left[\lim_{\Spaa{i|k}\to 0} P_{ik}^2(z)M_L (\WH p_i(z_{iI}), -\WH
P_{iI}(z_{iI}))\right]{M_R (\WH P_{iI}(z_{iI}), \WH
p_j(z_{iI}))\over P_{iI}^2(z)} \prod_{l=1}^{v+1} {w_l-z\over
w_l-z_{iI}}~~~\label{Pik-bcfw-z-1} \eea
where the sum is over all sets $I$ containing $k$ and at least one of other particles.
Using the factorization limit of $M_L$
\bean \lim_{\Spaa{i|k}\to 0} P_{ik}^2(z_{iI})M_L(\WH i(z_{iI}), k,
..., -\WH P_{ik\a}(z_{iI})) = M_3^{(a)}(\WH i(z_{iI}), k, -\WH
P_{\WH i k}(z_{iI})) M(\WH P_{\WH i k}(z_{iI}),..., -\WH
P_{ik\a}(z_{iI}))\eean
where $I= \a\bigcup k$ and  the notation $z_{iI}$ emphasizes that momenta are taken at the shifted value $z=z_{iI}$.
Putting this back to (\ref{Pik-bcfw-z-1}) we have
\bea R_{ik}(z)=\sum_{I} M_3^{(a)}(\WH i(z_{iI}), k, -\WH P_{\WH i
k}(z_{iI}))\left\{ {\Spbb{i(z)|k}\over \Spbb{i(z_{iI})|k}} M(\WH
P_{\WH i k},..., -P_{ik\a}){1\over P_{iI}^2(z)} M_R (\WH P_{iI}, \WH
p_j)\prod_{l=1}^{v+1} {w_l-z\over
w_l-z_{iI}}\right\}~~~\label{Pik-bcfw-z-2} \eea
which should be equal to the $D_{ik}$ in (\ref{Pik-A-z-1}).
Comparing these two functions of $z$, we now try to find
(1) the number of roots and how it changes under the limit;
(2) the values of roots and their behavior under the limit.

To make calculations clear, we will take three steps:
\begin{itemize}

\item {\bf Step One:} One difference between $R_{ik}$ and $D_{ik}$ is that the $M_3^a$ inside $R_{ik}(z)$ depends on cuts $I$,
while the $M_3^a$ inside $D_{ik}(z)$ is universal.
Thus using (\ref{3-a}) we can rewrite
\bean
M_3^a( i(z_{iI})^{h_i}, k^{h_k},-P_{ik}^{-h_{ik}}(z_{iI}))=
\left({\Spbb{i(z_{iI})|k}\over
\Spbb{i(z)|k}}\right)^{h_i+h_k-h_{ik}}M_3^a( i(z)^{h_i},
k^{h_k},-P_{ik}^{-h_{ik}}(z))
\eean
and
\bean {R_{ik}(z)\over M_3^a( i(z)^{h_i},
k^{h_k},-P_{ik}^{-h_{ik}}(z))}=\sum_{k\in I}
\left\{\left({\Spbb{i(z_{iI})|k}\over
\Spbb{i(z)|k}}\right)^{h_i+h_k-h_{ik}-1}  {M(\WH P_{\WH i k},...,
-P_{ik\a})  M_R (\WH P_{iI}, \WH p_j)\over
P_{iI}^2(z)}\prod_{l=1}^{v+1} {w_l-z\over
w_l-z_{iI}}\right\}~.~~~\label{Pik-bcfw-z-3}\eean
Identifying  $R_{ik}(z)$ with $D_{ik}(z)$ we obtain a consistent condition
\bea & & M_{n-1} (P_{ik}^{h_{ik}}(z),..,j(z),..)\nn & = & \sum_{I}
\left\{\left({\Spbb{i(z_{iI})|k}\over
\Spbb{i(z)|k}}\right)^{h_i+h_k-h_{ik}-1}  {M(\WH P_{\WH i k},...,
-\WH P_{iI}) M_R (\WH P_{iI}, \WH p_j)\over
P_{iI}^2(z)}\prod_{l=1}^{v+1} {w_l-z\over
w_l-z_{iI}}\right\}~~~\label{Pik-Comp-z-1} \eea
where
\bea P_{ik}(z) & = & \ket{i} \left( \bket{i}-z\bket{j}+
\bket{k}{\Spaa{k|\mu}\over \Spaa{i|\mu}}\right), ~~~p_j(z)=
(\ket{j}+z\ket{i})\bket{j},~~~p_i(z)=\ket{i}(\bket{i}-z\bket{j})~.~~\label{Pik-z-para}\eea

\item {\bf Step Two:} With a little calculation,
one sees that $M_{n-1} (P_{ik}^{h_{ik}}(z),..,j(z),..)$ in (\ref{Pik-Comp-z-1}) is
the $z$-dependent amplitude with the BCFW-deformation $\Spba{P_{ik}|j}$
(since under the limit $P_{ik}$ is null, its spinor and anti-spinor components are well defined),
so it can be expanded
\bea M_{n-1} (P_{ik}^{h_{ik}}(z),..,j(z),..)=\sum_{ I} {M(\WH P_{\WH
i k},..., -\WH P_{iI})  M_R (\WH P_{iI}, \WH p_j)\over
P_{iI}^2(z)}\prod_{l=1}^{\W v+1} {w_l-z\over
w_l-z_{iI}}~~~\label{Pik-Comp-z-2}\eea
where the number of roots is $\W v+1$ now. Comparing
(\ref{Pik-Comp-z-2}) with (\ref{Pik-Comp-z-1}), we observe that\footnote{Notice that  $h_i+h_k-h_{ik}$
is always a non-negative integer.}:
    \begin{itemize}

    \item (a-1) When $h_i+h_k-h_{ik}-1=0$, as in the case of gauge
    theory,  the number of roots of $n$-point amplitude is
    the same as the number of roots of $n-1$-point amplitude. In
    other words, the number of roots is independent of number
    of particles.

    \item (a-2) When $h_i+h_k-h_{ik}-1=1$, which includes the case of gravity theory,
    there are two possibilities.
    In one case, $n$-point amplitude has one more root than $n-1$-point amplitude.
    In the other case, they have the same number of roots,
    but there is a nontrivial cancelations.

    For gravity theory, the second possibility is realized and the nontrivial cancelation has been discussed carefully in \cite{Schuster:2008nh}
    (eq.(75)) as the bonus relation \cite{Benincasa:2007qj, ArkaniHamed:2008yf}.

    \end{itemize}

\item  {\bf Step Three:} The consistent condition under the factorization limit $\Spaa{i|k}\to 0$ is summarized as
\bea & & \sum_{I} \left\{\left({\Spbb{i(z_{iI})|k}\over
\Spbb{i(z)|k}}\right)^{h_i+h_k-h_{ik}-1}  {M(\WH P_{\WH i k},...,
-\WH P_{iI}) M_R (\WH P_{iI}, \WH p_j)\over
P_{iI}^2(z)}\prod_{l=1}^{v+1} {w_l-z\over w_l-z_{iI}}\right\}\nn
& = & \sum_{ I} {M(\WH P_{\WH i k},..., -\WH P_{iI})  M_R (\WH
P_{iI}, \WH p_j)\over P_{iI}^2(z)}\prod_{l=1}^{\W v+1} {w_l-z\over
w_l-z_{iI}}~.~~~~\label{Spaa-ik-con}\eea
If  one has only one term in the sum, which may happen for low point amplitudes, we will arrive at
\bea  \left({\Spbb{i(z_{iI})|k}\over
\Spbb{i(z)|k}}\right)^{h_i+h_k-h_{ik}-1}  \prod_{l=1}^{v+1}
{w_l-z\over w_l-z_{iI}}= \prod_{l=1}^{\W v+1} {w_l-z\over
w_l-z_{iI}}~.~~~\label{Pik-check}\eea
If there are more than one term in the sum, (\ref{Pik-check})
could be true for each term but unlikely.

\end{itemize}
%

\subsubsection*{Potentially extra singularities at $\Spaa{i|k}=0$}

Above discussions have a small loop hole, i.e., there are some
potential contributions we have overlooked. For example, for $n=4$,
there is no term in the summation in (\ref{Pik-bcfw-z-1}) and there
must be some place to provide the needed contribution\footnote{
Related singular behavior is that if $P_{12}^2\to 0$, then
$P_{34}^2\to 0$, thus we will have $\ket{1}\sim \ket{2}$ and
$\bket{3}\sim \bket{4}$ at same time.}. The potential contribution
comes from following term
\bea T_{jk}(z)= M_3^{(a)}(\WH j, k,-\WH P_{jk}^{-h_{jk}}){1\over
(\Spaa{j|k}+z\Spaa{i|k})\Spbb{k|j}}M_{n-1}(\WH P_{jk}^{h_{jk}}, \WH
i..)\prod_{l=1}^{v+1} {w_l-z\over w_l-z_{jk}}~~~~\label{Pik-jk-z-1}
\eea
where\footnote{Notice the unusual definition of the spinor and anti-spinor components of $\WH P_{jk}$.
The reason for this definition will be clear from later discussions.}
\bea z_{jk}=-{\Spaa{j|k}\over \Spaa{i|k}},~~~\bket{\WH i}=
{\ket{P_{ij}|k}\over \Spaa{i|k}},~~~~\ket{\WH
j}=\ket{k}{\Spaa{i|j}\over \Spaa{i|k}},~~~\WH P_{jk}
=\left({\Spaa{i|j}\over
\Spaa{i|k}}\ket{k}\right)\left(\bket{j}+{\Spaa{i|k}\over
\Spaa{i|j}}\bket{k}\right)~.~~~~\label{Pik-jk-pare-1} \eea
Notice that under the limit $\Spaa{i|k}\to 0$, $z_{jk}, \ket{\WH j}$ and $\bket{\WH i}$ are all going to infinity.
The $M_3^{(a)}$ part now becomes
\bea M_3^{(a)}(\WH j, k,-\WH P_{jk}^{-h_{jk}})\sim
\Spbb{j|k}^{h_k+h_j-h_{jk}}(-)^{-2h_{jk}}\left({\Spaa{i|k}\over
\Spaa{i|j}}\right)^{-h_k-h_{jk}+h_j}~. \eea
The understanding of  $M_{n-1}(\WH P_{jk}^{h_{jk}}, \WH i,...)$ part
can be given as following. Define the ``initial momenta"
\bea p_i^{init}= \ket{i}\left( \bket{i}+{\Spaa{j|k}\over \Spaa{j|i}}
\bket{k}\right),~~~P_{jk}^{init}= \ket{j}\left(
\bket{j}+{\Spaa{i|k}\over \Spaa{i|j}} \bket{k}\right),
~~~P_{jk}^{init}+p_i^{init}=p_i+p_j+p_k\eea
and use them to do the  $\Spba{i^{init}| P_{jk}^{init}}$-deformation
\bea \bket{p_i^{init}}\to \bket{p_i^{init}}-z
\bket{P_{jk}^{init}},~~~~~\ket{P_{jk}^{init}}\to
\ket{P_{jk}^{init}}+z \ket{p_{i}^{init}}~,\eea
then it is easy to see that  when we set $z=-{\Spaa{j|k}\over
\Spaa{i|k}}$, we produce right spinor variables
(\ref{Pik-jk-pare-1}).

In (\ref{Pik-jk-z-1}),
one finds three contributions for the overall power of factor  $\Spaa{i|k}$:
(1) $M_3$ gives a power of $-h_k-h_{jk}+h_j$;
(2) If $M_{n-1} \sim z^t$ at the infinity under the deformation $\Spba{i^{h_i}| P_{jk}^{h_{jk}}}$,
there is factor $\Spaa{i|k}^{-t}$;
(3) The $\prod_l {(w_l-z)/(w_l-z_{jk})}$ could give
another power of $\Spaa{i|k}^{\nu}$ with $\nu\leq v+1$.
This happens when root $w_l$ is finite under the limit,
thus ${(w_l-z)/ (w_l-z_{jk})}\to {\Spaa{i|k}(w_l-z)/\Spaa{j|k}}$.
Collecting all factors together,  we have finally
\bea \Spaa{i|k}^{n_{ik}}\equiv
\Spaa{i|k}^{-h_k-h_{jk}+h_j-t+\nu}~~~\label{Potential-index}\eea
If $n_{ik}$ is non-negative, there is no contribution in the limit $\Spaa{i|k}\to 0$.
If $n_{ik}$ is negative, it does give non-zero contribution.
To have a finite factorization limit, one needs $n_{ik}=-1$,
from which $\nu$ could be obtained.
However,  without a general expression for $M_{n-1}$,
detailed informations can be only inferred in explicit example.

\subsubsection{The pole $\Spbb{j|k}=0$ from the cut $P_{jJ}$ with $k\in J$}

This part parallels to the discussion of $\Spaa{i|k}=0$.
The factorization limit from (\ref{Mz-new}) gives
\bea \lim_{\Spbb{j|k}\to 0} P_{jk}^2(z)M_n(z)& \equiv & R_{jk}(z)
\nn & = & \sum_{k\in J} M_3^{(h)}(\WH j(z_{jJ}), k, -\WH P_{\WH j
k}(z_{jJ}))\left\{ {\Spaa{j(z)|k}\over \Spaa{j(z_{jJ})|k}} {1\over
P_{jJ}^2(z)} M(\WH P_{\WH j k}(z_{jJ}),..., -\WH
P_{jJ}(z_{jJ}))\right. \nn & & \left. M_R (\WH P_{jJ}(z_{jJ}), \WH
p_j(z_{jJ}))\prod_{l=1}^{v+1} {w_l-z\over
w_l-z_{jJ}}\right\}~~~\label{Pjk-bcfw-z-2}\eea
where the  sum is over all $J$ containing $k$ and at least another particle.
Consistent conditions can again be obtained by comparing $R_{jk}$ with $D_{jk}$ from (\ref{Pjk-A-z-1}).
\begin{itemize}

\item {\bf Step One:}    Rewrite
\bea M_3^h( j(z_{jI})^{h_j}, k^{h_k},-P_{jk}^{-h_{jk}}(z_{jI}))=
\left({\Spaa{j(z_{jI})|k}\over
\Spaa{j(z)|k}}\right)^{-(h_j+h_k-h_{jk})}M_3^h( j(z)^{h_j},
k^{h_k},-P_{jk}^{-h_{jk}}(z))\eea
and compare with the $D_{jk}(z)$ in (\ref{Pjk-A-z-1}), one has the following consistent condition
\bea & & M_{n-1}
(P_{jk}^{h_{jk}}(z),..,i(z),..)~~~\label{Pjk-Comp-z-1} \\& =
&\left({\Spaa{j+z_{jJ} i|k}\over \Spaa{j+z
i|k}}\right)^{-(h_j+h_k-h_{jk})-1} {M(\WH P_{\WH j k}(z_{jJ}),...,
-\WH P_{jJ}(z_{jJ})) M_R (\WH P_{jJ}(z_{jJ}), \WH p_j(z_{jJ}))\over
P_{jJ}^2(z)}\prod_{l=1}^{v+1} {w_l-z\over w_l-z_{jJ}}\nonumber  \eea
here
\bea P_{jk}(z) & = &  \left( \ket{j}+z\ket{i}+
\ket{k}{\Spbb{k|\mu}\over \Spbb{j|\mu}}\right)\bket{j}, ~~~p_j(z)=
(\ket{j}+z\ket{i})\bket{j},~~~p_i(z)=\ket{i}(\bket{i}-z\bket{j})~.~~\label{Pjk-z-para}\eea

\item  {\bf Step Two:} $M_{n-1} (P_{jk}^{h_{jk}}(z),..,i(z),..)$ is  obtained by the $\Spba{i|P_{jk}}$-deformation
and can be expanded as
\bea M_{n-1} (P_{jk}^{h_{jk}}(z),..,i(z),..)=\sum_{ J}{M(\WH P_{\WH
j k}(z_{jJ}),..., -\WH P_{jJ}(z_{jJ})) M_R (\WH P_{jJ}(z_{jJ}), \WH
p_j(z_{jJ}))\over P_{jJ}^2(z)} \prod_{l=1}^{\W v+1} {w_l-z\over
w_l-z_{jJ}}~~~\label{Pjk-Comp-z-2} \eea
Now the number of zero is $\W v+1$.
If $h_j+h_k-h_{jk}+1=0$, the number of roots of $n$ point amplitude is the same as that of $n-1$ point amplitude.
If $h_j+h_k-h_{jk}+1\leq -1$, we should study carefully about how the match is realized.

\item {\bf Step Three:} If the sum over $J$ has only one term, we will have
\bea \left({\Spaa{j+z_{jJ} i|k}\over \Spaa{j+z
i|k}}\right)^{-(h_j+h_k-h_{jk})-1}\prod_{l=1}^{v+1} {w_l-z\over
w_l-z_{jJ}}= \prod_{l=1}^{\W v+1} {w_l-z\over
w_l-z_{jJ}}~.~~~~\label{Pjk-check} \eea
Again we cannot  be certain whether (\ref{Pjk-check}) is true for every possible cut.

\end{itemize}
%

\subsubsection*{Potentially extra singularities at pole $\Spbb{j|k}=0$}

A possible contribution comes from the following term\footnote {In
fact, if the summation over $J$ is empty, this term must contribute
to get consistent result.}
\bea T_{ik}(z)= M_3^{(h)}(\WH i, k,-\WH P_{ik}^{-h_{ik}}){1\over
s_{ik}(z)} M_{n-1}(\WH P_{ik}^{h_{ik}}, \WH j,..)\prod_{l=1}^{v+1}
{w_l-z\over w_l-z_{ik}}~~~~\label{Pjk-ik-z-1} \eea
with definitions of various quantities as
\bea z_{ik}={\Spbb{i|k}\over \Spbb{j|k}},~~~\bket{\WH i}=
\bket{k}{\Spbb{j|i}\over \Spbb{j|k}},~~~~\ket{\WH
j}=\ket{j}+{\Spbb{i|k}\over \Spbb{j|k}}\ket{i},~~~\WH P_{ik} =\left(
\ket{k}{\Spbb{j|k}\over \Spbb{j|i}}+
\ket{i}\right)\left(\bket{k}{\Spbb{j|i}\over \Spbb{j|k}}
\right)~.~~~\label{Pjk-ik-pare-1} \eea
Under the limit $\Spbb{j|k}\to 0$, $z_{ik}, \ket{\WH j}$ and $\bket{\WH i}$ are all going to infinity.
As in the case of $\Spaa{i|k}$, we now count the singularity power.
The first factor comes from
\bea
M_3^h (i^{h_i}(z), k^{h_k},
-P_{ik}^{-h_{ik}}(z))=(-)^{2h_{jk}} \Spaa{ i|k}^{-(h_i+h_k-h_{ik})}
\left( {\Spbb{j|k}\over \Spbb{j|i}}\right)^{h_k+h_{ik}-h_i}~.
\eea
The second one comes from the infinity behavior $z^{t}$ of
 $M_{n-1}(\WH P_{ik}^{h_{ik}}, \WH
j,...)$ with ``initial momenta"
\bea
p_j^{init}= \left(\ket{j}-{\Spbb{i|k}\over
\Spbb{j|i}}\ket{k}\right)\bket{j},~~~P_{ik}^{init}=\left(
\ket{k}{\Spbb{j|k}\over \Spbb{j|i}}+ \ket{i}\right)\bket{i},
~~~P_{ik}^{init}+p_j^{init}=p_i+p_j+p_k
\eea
and  the $\Spba{ P_{ik}^{init}|j^{init}}$ deformation\footnote{
Setting $z={\Spbb{i|k}/ \Spbb{j|k}}$, we reproduce right spinor variables in (\ref{Pjk-ik-pare-1}).}
\bea \ket{p_j^{init}}\to \ket{p_j^{init}}+z
\ket{P_{jk}^{init}},~~~~~\bket{P_{jk}^{init}}\to
\bket{P_{jk}^{init}}-z \bket{p_{j}^{init}}~.\eea
The third one comes from $\prod_l {(w_l-z)/(w_l-z_{jk})}$  of power
$\nu\leq v+1$.  Collecting these together, the final power is
\bea \Spbb{j|k}^{n_{jk}}\equiv
\Spbb{j|k}^{h_k+h_{ik}-h_i-t+\nu}~~~\label{Potential-index-1} \eea
If $n_{jk}$ is non-negative, there is no contribution in the limit of $\Spbb{j|k}\to 0$.
If it is negative, it does give non-zero contribution and
$\nu$ can be determined by requiring $h_k+h_{ik}-h_i-t+\nu=-1$.

\subsection{Poles without $z$-dependence}

Among poles without $z$-dependence, two-particle pole $P_{ij}=0$ will be particularly important.
The general pole $P_\a$ here appears in cuts $T_{iI}$ (\ref{T-iI}) with  $\a\subset I$ or $\a\subset \O {i,j,I}$.
Under the limit $P_{\a}^2\to 0$, we have
\bea
& & \lim_{P_{\a}^2\to 0} P_{\a}^2 M_n(z)\nn
&= &  \lim_{P_{\a}^2\to 0} P_{\a}^2\left\{\sum_{I} M_L (\WH p_i,
-\WH P_{iI}){1\over P_{iI}^2(z)} M_R (\WH P_{iI}, \WH
p_j)\prod_{l=1}^{v+1}{ w_l-z\over w_l-z_{iI}}\right\}\nn
& = & \sum_{\a\in I} M(\{\a\}, -P_{\a}) \W M_{L} (\WH p_i, -\WH
P_{iI}, P_{\a}){1\over P_{iI}^2(z)} M_R (\WH P_{iI}, \WH
p_j)\prod_{l=1}^{v+1}{ w_l-z\over w_l-z_{iI}}\nn
 & & +\sum_{\a \in \O{i,j,I}} M_L
(\WH p_i, -\WH P_{iI}){1\over P_{iI}^2(z)} M_R (\WH P_{iI},
\WH p_j, P_\a) M(\{\a\}, -P_{\a})\prod_{l=1}^{v+1}{ w_l-z\over
w_l-z_{iI}}\nn
& \equiv & M(\{\a\}, -P_{\a})
M_{n-\{\a\}}(z)~~~~\label{P-alpha-red}\eea
The number of roots of $M_{n-\{\a\}}$ should be the same or less than that in $M_n$.
They are the same only if no $w_l\to \infty$ when $P_\a^2\to 0$.
Therefore, values of these non-divergent roots $w_l$ under the limit $P_\a^2\to 0$
could be read out from lower point amplitudes $M_{n-\{\a\}}(z)$ with the same $z$-deformation.

Now some remarks on (\ref{P-alpha-red}) about the two particle channel $P_{k_1 k_2}$.
The term $M_L(\WH i,k_1, k_2, -\WH P_{ik_1 k_2})$  as given in (\ref{general-bcfw-z})
will give zero contribution under $\Spbb{k_1|k_2}\to 0$.
Since $z={\Spbb{k_1|i}/ \Spbb{k_1|j}}$ under the limit,  so
\bean
\WH{\W\la}_i& = & \W\la_i -z\W\la_j= \W\la_{k_1}
{\Spbb{i|j}\over \Spbb{k_1|j}}
\eean
and
\bean
\W\la_{k_1}\sim \W\la_{k_2}\sim \W\la_{P_{k_1k_2}}\sim
\WH{\W\la}_{i}\sim \WH{\W\la}_{P_{ik_1 k_2}}.
\eean
Thus
\bea \lim_{\Spbb{k_1|k_2}\to 0}P^2_{k_1 k_2}M(\WH i,k_1, k_2, -\WH
P_{ik_1 k_2}) & \to & M_3(k_1, k_2, -P_{k_1k_2}) M_3(P_{k_1
k_2},-\WH P_{ik_1 k_2}, \WH i)\to 0~~~\label{P2-anti-1} \eea
because the product in (\ref{P2-anti-1}) must be $M_3^a\times M_3^h$ and $M_3^a=0$.
This null contribution was explained in \cite{Schuster:2008nh},
while $M_3(\WH i, k_1, -\WH P_{ik_1}) M_{n-1}(-\WH P_{ik_1}, k_2, \WH j,...)$
(plus the term with $k_1\leftrightarrow k_2$) provides an extra contribution.
Similar subtleties arise when $\Spaa{k_1|k_2}\to 0$ and more can be found in \cite{Schuster:2008nh}.

Now we turn to the pole at $P_{ij}=0$.
It does not appear explicitly in the recursion relation (\ref{general-bcfw-z}).
Since $P_{ij}(z)=P_{ij}$ for all $z$, factorization limits exists for all values of  $z$:
\bea \lim_{\Spbb{i|j}\to 0} P_{ij}^2 M_n & \to &
M_3^{h}(i(z),j(z),-P_{ij}^{-h_{ij}})
M_{n-1}(P_{ij}^{+h_{ij}},...)~~~\label{Pij-holo} \eea
and
\bea \lim_{\Spaa{i|j}\to 0} P_{ij}^2 M_n & \to &
M_3^{a}(i(z),j(z),-P_{ij}^{-h_{ij}})
M_{n-1}(P_{ij}^{+h_{ij}},...)~.~~~\label{Pij-anti} \eea
Depending on the helicity configuration $h_i, h_j$, both limits can
be nontrivial or one of them be trivial, for example, if $h_i,h_j$
make $A^{h}(i,j,-P_{ij})=0$ no matter what the helicity of $P_{ij}$
is. The two singularities $\Spaa{i|j}\to 0$ and $\Spbb{i|j}\to 0$ do
not come from spurious poles $\Spab{i|P_{i\a} (P_{i\a}-P_{iI})
P_{iI}|j}$ or $\Spab{i|P_{iI} (P_{iI}-P_{i\a}) P_{i\a}|j}$, but
 the {\bf soft
limit} of $\WH p_i$ or $\WH p_j$ in  following two types of cut
contributions (other terms do not give soft limit and wanted
singularities)
\bea
T_{ik}(z)&= &M_3^{h} (\WH i^{h_i}, k^{h_k}, -\WH
P_{ik}^{-h_{ik}}){1\over s_{ik}(z)} M_{n-1}(\WH P_{ik}^{h_{ik}}, \WH
j^{h_j},...)\prod_l {w_l-z\over w_l-z_{ik}} \nn
T_{jk}(z)& = & M_3^{a} (\WH j^{h_j}, k^{h_k}, -\WH P_{jk}^{-h_{jk}})
{1\over s_{jk}(z)} M_{n-1}(\WH P_{jk}^{h_{jk}}, \WH
i^{h_i},...)\prod_l {w_l-z\over w_l-z_{jk}}~. ~~~~\label{Pole-ij-1}
\eea
In $T_{ik}(z)$
\bea z_{ik}={\Spbb{k|i}\over \Spbb{k|j}},~~~\bket{\WH i}=\W\la_k
{\Spbb{i|j}\over \Spbb{k|j}},~~~\WH
P_{ik}=\W\la_k\left(\la_i{\Spbb{i|j}\over \Spbb{k|j}}+ \la_k
\right),~~~\WH p_j= \bket{j}\left( \ket{j}+ {\Spbb{k|i}\over
\Spbb{k|j}}\ket{i}\right)~~~~\label{Tik-1} \eea
which gives $\bket{\WH i}\to 0$ under the limit $\Spbb{i|j}\to 0$.
In $T_{jk}(z)$,
\bea z_{jk}=-{\Spaa{j|k}\over \Spaa{i|k}},~~~\bket{\WH
i}=\bket{i}+{\Spaa{j|k}\over \Spaa{i|k}}\bket{j}, ~~~\ket{\WH
j}=\ket{k} {\Spaa{i|j}\over \Spaa{i|k}},~~~~\WH P_{jk}=\ket{k}
\left( \bket{j}{\Spaa{i|j}\over \Spaa{i|k}}+\bket{k}
\right)~~~~\label{Tjk-1} \eea
thus  $\ket{\WH j}\to 0$ under the limit $\Spaa{i|j}\to 0$.

\subsubsection{The $\Spbb{i|j}\to 0$ limit}

We now compare the factorization limit of $T_{ik}(z)$ with (\ref{Pij-holo}).
The $z$-independent part of  $M_3$ in $T_{ik}(z)$ is
\bea
M_3^{h} (\WH i, k, -\WH P_{ik}^{-h_{ik}})\sim (-)^{2 h_{ik}}
\Spaa{i|k}^{-(h_i+h_k- h_{ik})} \left( {\Spbb{i|j}\over
\Spbb{k|j}}\right)^{h_i+h_{ik}-h_k}\sim (-)^{2 h_{ik}}
\Spaa{i|k}^{\delta_{ik}^h} \left( {\Spbb{i|j}\over
\Spbb{k|j}}\right)^{2h_i+\delta_{ik}^h}
\eea
where  we have used (\ref{Tik-1}) and defined
\bea \delta_{ik}^h= -(h_i+h_k- h_{ik})\geq
0~.~~~~\label{delta-ik-h}\eea
The $z$-dependent part of $M_3$ in (\ref{Pij-holo}) is
\bea M_3^h(i(z),j(z),-P_{ij}^{-h_{ij}})\sim
(-)^{2h_{ij}}\Spaa{i|j}^{-(h_i+h_j-h_{ij})} \left({\Spbb{\mu|i}\over
\Spbb{\mu|j}} -z\right)^{h_i+h_{ij}-h_j}\sim
(-)^{2h_{ij}}\Spaa{i|j}^{\delta_{ij}^h} \left({\Spbb{\mu|i}\over
\Spbb{\mu|j}} -z\right)^{2h_i+\delta_{ij}^h}\eea
where we have used
\bea P_{ij}=\left(\ket{j}+\ket{i}{\Spbb{\mu|i}\over
\Spbb{\mu|j}}\right) \bket{j}~~~\label{Pij-anti-1}\eea
and defined
\bea  \delta_{ij}^h=-(h_i+h_j-h_{ij})\geq 0.~~~\label{delta-ij-h}
\eea

Now to the $M_{n-1}$ part of $T_{ik}$.
Under the limit $\Spbb{i|j}\to 0$, $  \WH p_j\to p_i+p_j=P_{ij}$ and $\WH P_{ik}\to p_k$,
the $M_{n-1}$ in $T_{ik}$ becomes $ M_{n-1}(p_k^{h_{ik}}, P_{ij}^{h_j},...)$.
To link it with the $M_{n-1}(P_{ij}^{h_{ij}}, p_k^{h_k},...)$  in (\ref{Pij-holo}), we define
\bea M_{n-1}(p_k^{h_{ik}}, P_{ij}^{h_j},...)= {\cal H}_{n-1}^{(ijk)}
M_{n-1}(P_{ij}^{h_{ij}}, p_k^{h_k},...)~~~\label{H-def-holo} \eea
where the function ${\cal H}_{n-1}^{(ijk)}$  is  $z$-independet.

Comparing (\ref{Pij-holo}) with $\sum_k P_{ij}^2 T_{ik}(z)$, one has
\bean & & \sum_k (-)^{2 h_{ik}} \Spaa{i|k}^{\delta_{ik}^h} \left(
{\Spbb{i|j}\over \Spbb{k|j}}\right)^{2h_i+\delta_{ik}^h}
{s_{ij}\over s_{ik}(z)}M_{n-1}(p_k^{h_{ik}}, P_{ij}^{h_j},...)\prod_l
{w_l-z\over w_l-z_{ik}} \nn
& = & (-)^{2h_{ij}}\Spaa{i|j}^{\delta_{ij}^h}
\left({\Spbb{\mu|i}\over \Spbb{\mu|j}}
-z\right)^{2h_i+\delta_{ij}^h}M_{n-1}(P_{ij}^{h_{ij}},
p_k^{h_k},...)\eean
which can be simplified to (the sign comes from
possible different color ordering)
\bea \pm 1&= & \sum_k (-)^{2 h_{ik}-2h_{ij}} {
\Spaa{i|k}^{\delta_{ik}^h} \over \Spaa{i|j}^{\delta_{ij}^h}}\left(
{\Spbb{i|j}\over
\Spbb{k|j}}\right)^{2h_i+\delta_{ik}^h}\left({\Spbb{\mu|i}\over
\Spbb{\mu|j}}-z \right)^{-(2h_i+\delta_{ij}^h)} {s_{ij}\over
s_{ik}(z)}{\cal H}_{n-1}^{(ijk)}\prod_l {w_l-z\over
w_l-z_{ik}}~.~~~\label{Pij-Cond-holo} \eea
Here are some points on (\ref{Pij-Cond-holo}):
    \begin{itemize}

    \item {\bf $z$-dependence:} Note $s_{ik}(z)=\Spaa{i|k}\Spbb{k|j}\left( {\Spbb{k|i}/\Spbb{k|j}}-z\right)$
    and ${\Spbb{k|i}/ \Spbb{k|j}}={\Spbb{\mu|i}/\Spbb{\mu|j}}$ for any $\mu$ under the limit.
    To make the right-handed side of (\ref{Pij-Cond-holo}) $z$-independent,
     the number of finite $w_l$ must be
    \bea N_{zero}^h= 1+ 2h_i+ \delta_{ij}^h~.~~~~\label{Nzero-Pij-h}
    \eea
For each $k$, the $z$-dependence of the factors
    $\left( {\Spbb{k|i}/ \Spbb{k|j}}-z\right)^{-1} \left({\Spbb{\mu|i}/\Spbb{\mu|j}}-z \right)^{-(2h_i+\delta_{ij}^h)}\prod_l(w_l-z)$ are the same, so they can be pulled out uniformly through the summation.

    \item {\bf Universal behavior}: A consequence of the above $z$-dependence is that values of roots will be
     \bea
     w_l\to {\Spbb{t_l|i}\over
\Spbb{t_l|j}}~~~~\label{Nero-form-Pij-h}
     \eea
    under the limit $\Spbb{i|j}\to 0$.

    \item {\bf Further simplification:} Using (\ref{Nero-form-Pij-h}) we have
    $ w_l-z_{ik}
    = \Spbb{t_l|k}\Spbb{i|j}/\Spbb{t_l|j}\Spbb{k|j}$,
    thus (\ref{Pij-Cond-holo}) can be further simplified to
    \bea \pm 1&= & \sum_k (-)^{2 h_{ik}-2h_{ij}} {
\Spaa{i|k}^{\delta_{ik}^h} \over \Spaa{i|j}^{\delta_{ij}^h}}\left(
{\Spbb{i|j}\over \Spbb{k|j}}\right)^{2h_i+\delta_{ik}^h}
{s_{ij}\over s_{ik}}{\cal H}_{n-1}^{(ijk)}\prod_{l=1}^{N_{zero}^h}
\frac{\Spbb{t_l|j}\Spbb{k|j}}{\Spbb{t_l|k}\Spbb{i|j}}~.~~~\label{Pij-Cond-holo-1}\eea

    \item {\bf A special case:}
    In cases such as pure gauge or gravity theory, $\delta_{ik}^h=\delta_{ij}^h$
    (though we cannot assume so in general).
    Under these circumstances, we can simplify further
    \bea
    \pm 1&= & \sum_k  {
\Spaa{i|k}^{\delta_{ik}^h} \over \Spaa{i|j}^{\delta_{ij}^h}}
{\Spaa{j|i}\Spbb{k|j}\over s_{ik}}{\cal
H}_{n-1}^{(ijk)}\prod_{l=1}^{N_{zero}^h}
\frac{\Spbb{t_l|j}}{\Spbb{t_l|k}}\nn & = &\sum_k  \left({ \Spaa{i|k}
\over \Spaa{i|j}}\right)^{\delta-1} {\Spbb{k|j}\over
\Spbb{i|k}}{\cal H}_{n-1}^{(ijk)}\prod_{l=1}^{N_{zero}^h}
\frac{\Spbb{t_l|j}}{\Spbb{t_l|k}}~. ~~~\label{Pij-Cond-holo-2} \eea

    \end{itemize}
    %

\subsubsection{The $\Spaa{i|j}\to 0$ limit}

Now compare the factorization limit of $T_{jk}(z)$ with (\ref{Pij-anti}).
The discussion will be brief due to its similarity to the previous one.
The $z$-independent part of $M_3$ in $T_{jk}(z)$ is
\bea
M_3^{a} (\WH j, k, -\WH P_{jk}^{-h_{jk}})\sim (-)^{-2 h_{jk}}
\Spbb{j|k}^{(h_j+h_k- h_{jk})} \left( {\Spaa{i|j}\over
\Spaa{i|k}}\right)^{-h_j-h_{jk}+h_k}\sim (-)^{-2 h_{jk}}
\Spbb{j|k}^{\delta_{jk}^a} \left( {\Spaa{i|j}\over
\Spaa{i|k}}\right)^{-2h_j+\delta_{jk}^a}
\eea
where $ \delta_{jk}^a= (h_j+h_k- h_{jk})\geq 0$.
The $z$-independent part of $M_3$ in (\ref{Pij-holo}) is
\bea
M_3^a(i(z),j(z),-P_{ij}^{-h_{ij}})\sim
(-)^{-2h_{ij}}\Spbb{i|j}^{(h_i+h_j-h_{ij})} \left({\Spaa{\mu|j}\over
\Spaa{\mu|i}} +z\right)^{h_i-h_{ij}-h_j}\sim
(-)^{-2h_{ij}}\Spbb{i|j}^{\delta_{ij}^a} \left({\Spaa{\mu|j}\over
\Spaa{\mu|i}} +z\right)^{\delta_{ij}^a-2h_j}
\eea
where  $\delta_{ij}^a=(h_i+h_j-h_{ij})\geq 0$.
To link $ M_{n-1}(p_k^{h_{jk}}, P_{ij}^{h_i},...)$ in $T_{jk}$  with $M_{n-1}(P_{ij}^{h_{ij}}, p_k^{h_k},...)$ in (\ref{Pij-holo}),
we define
\bea M_{n-1}(p_k^{h_{jk}}, P_{ij}^{h_i},...)= {\cal \W
H}_{n-1}^{(ijk)} M_{n-1}(P_{ij}^{h_{ij}},
p_k^{h_k},...)~.~~~\label{H-def-anti} \eea
The  comparison of (\ref{Pij-holo}) with $\sum_k P_{ij}^2
T_{ik}(z)$ leads to following equation
\bea \pm 1&= & \sum_k (-)^{-2 h_{jk}+2h_{ij}}
{\Spbb{j|k}^{\delta_{jk}^a}\over \Spbb{i|j}^{\delta_{ij}^a}} \left(
{\Spaa{i|j}\over \Spaa{i|k}}\right)^{-2h_j+\delta_{jk}^a}
\left({\Spaa{\mu|j}\over \Spaa{\mu|i}}+z
\right)^{-(\delta_{ij}^a-2h_j)}{s_{ij}\over s_{jk}(z)}{\cal \W
H}_{n-1}^{(ijk)}\prod_l {w_l-z\over
w_l-z_{jk}}~~~\label{Pij-Cond-anti} \eea
from which one observes:
    \begin{itemize}

    \item {\bf  $z$-dependence:} To cancel the
    $z$-dependence on the right-handed side,  the number of finite $w_l$ is
    \bea N_{zero}^a= 1+\delta_{ij}^a-2h_j~.~~~\label{Nzero-Pij-a}\eea

    \item {\bf Universal behavior:} Values of root will be
    \bea w_l\to -{\Spaa{t_l|j}\over
\Spaa{t_l|i}}~~~\label{Nero-form-Pij-a}\eea
    under the limit $\Spaa{i|j}\to 0$.

    \item {\bf Further simplification:} Using
    $  w_l-z_{jk}= \Spaa{j|i}\Spaa{t_l|k}/ \Spaa{i|k}\Spaa{t_l|i}$, (\ref{Pij-Cond-anti})
    can be simplified further to
\bea \pm 1&= & \sum_k {\Spbb{j|k}^{\delta_{jk}^a}\over
\Spbb{i|j}^{\delta_{ij}^a}} \left( {\Spaa{i|j}\over
\Spaa{i|k}}\right)^{-2h_j+\delta_{jk}^a} {s_{ij}\over s_{jk}}{\cal
\W H}_{n-1}^{(ijk)}\prod_{l=1}^{N_{zero}^a}{\Spaa{i|k}\Spaa{t_l|i}
\over \Spaa{j|i}\Spaa{t_l|k}}~.~~~\label{Pij-Cond-anti-1} \eea

    \item {\bf A special case:} Assuming
$\delta_{jk}^a=\delta_{ij}^a$,  we have
\bea \pm 1&= & \sum_k \left({\Spbb{j|k}\over
\Spbb{i|j}}\right)^{\delta-1} {\Spaa{i|k}\over \Spaa{j|k}}{\cal \W
H}_{n-1}^{(ijk)}\prod_{l=1}^{N_{zero}^a}{\Spaa{t_l|i} \over
\Spaa{t_l|k}}~.~~~\label{Pij-Cond-anti-2} \eea

    \end{itemize}
    %

\section{Examples}

Listed in this section are examples \cite{Benincasa:2011kn} to
demonstrate previous general discussions about roots. We will show
(1) how the number  of roots behaves under various $z$-dependent
factorization limits; (2) how to infer values of roots from rational
function of $z$ under $z$-dependent factorization limits, when
possible; (3) finally, to show the limitation of our approach in the
case of six-gluon amplitudes.

\subsection{Example I-- MHV amplitudes}

We start with the simplest case,
the MHV amplitudes $M_n(-,+,-,+,..+,+)$ with deformation $\la_1\to \la_1+z\la_2$,
$\W\la_2\to \W\la_2-z\W \la_1$,  or the $\Spba{2|1}$-deformation.
There is only one pole and one gets from (\ref{Mz-new})
\bea
M_n(z)&= & M^a_3(n^+, \WH 1^-,- P^+) {1\over s_{n1}(z)} M(P^-, \WH 2^+, 3^-,...,(n-1)^+) \prod_{l=1}^{v+1} { w_l-z\over w_l- z_\a}\nn
& = & {-1\over \Spaa{1|2}\Spaa{2|3}...\Spaa{n|1}} \left(
{\Spaa{n|3}\Spaa{1|2}\over \Spaa{n|2}} \right)^4 {\Spaa{1|n}\over
\Spaa{1|n}+z\Spaa{2|n}}\prod_{l=1}^{v+1} {w_l-z\over w_l-
z_\a}~~~\label{MHV-wrong-z} \eea
with $z_\a=-\Spaa{1|n}/\Spaa{2|n}$.

\subsubsection{Poles without $z$-dependence }

The $M_{n-1}(z)$ part in (\ref{P-alpha-red}) is
$\Spaa{\WH1|3}^4/\Spaa{\WH1|2}...\Spaa{a-1|P_{a,a+1}}\Spaa{P_{a,a+1}|a+2}...\Spaa{n|\WH1}$.
The collinear limits  are\footnote
{There is no multiple-particle channel and only one nontrivial choice in the limit $\Spaa{a|a+1}\to 0$.}
$ \Spaa{a^+ |(a+1)^+}\to 0$ with $4\leq a\leq n-1$.
Under this limit and with the same deformation $\Spba{2|1}$,
we find a root $w_l=-{\Spaa{1|3}/\Spaa{2|3}}$ of multiplicity 4.
In the original amplitude without taking the limit, we should have
\bea
w_l= -{\Spaa{1|3}\over \Spaa{2|3}}\left( 1+f_l\right)
\eea
where $f_l$ should be constrained by several physical requirements:
(1) $f_l$ should have a factor $\Spaa{a|a+1}$ to give the root under the limit;
(2) $f_l$ should be helicity neutral for all particles,
      of either the form $s_{a,a+1}$ or the combination ${\Spaa{a|a+1}\Spaa{t|s}/\Spaa{a|s}\Spaa{a+1|t}}$
      of spinors $\la_t,\la_s$;
(3) $f_l$ should be dimensionless;
(4) $f_l$ should be consistent with all different choices $\Spaa{a|a+1}\to 0$;
(5) there must be no un-physical pole from $\prod_{l=1}^{v+1} { w_l/ (w_l- z_\a)}$ when $f_l$ is included.
Consistent with these requirements, there is simple solutions $f_l=0$, for $l=1,2,3,4$.

\subsubsection{Poles with $z$-dependence}

Here $s_{n1}=0$ and $s_{23}=0$ are pole of $z$-dependence.
For $s_{n1}=0$, the limit $\Spaa{n|1}\to 0$ is automatically satisfied by (\ref{Mz-new}),
while $M_{n-1}(P_{n1}^+, 2^+, 3^-,....)=0$ under the limit $\Spbb{n|1}\to0 $.
These are trivial.
For $s_{23}$, the $\Spbb{2|3}\to 0$ limit is trivial and we will focus on the $\Spaa{2|3}\to 0$ limit.
A new feature arises under this limit.
The true root $w_l=-{\Spaa{1|3}/\Spaa{2|3}}\to \infty$ and ${(w_l-z)/(w_l-z_{n1})}\to 1$,
so the degree of $z$ is reduced in the combination.
Let's see how this happen.

The factorization limit from (\ref{MHV-wrong-z}) is
\bean \lim_{\Spaa{2|3}\to 0}
(\Spbb{3|2}-z\Spbb{3|1})\Spaa{2|3}{-1\over
\Spaa{1|2}\Spaa{2|3}...\Spaa{n|1}} \left( {\Spaa{n|3}\Spaa{1|2}\over
\Spaa{n|2}} \right)^4 {\Spaa{n|1}\over
\Spaa{n|1}+z\Spaa{n|2}}\prod_{l=1}^{v+1} { w_l-z\over w_l- z_\a}\eean
while the direct factorization limit is
\bean  \Spbb{2-z 1|3} \left( {\Spaa{3|\mu}\over
\Spaa{2|\mu}}\right)^3 {\Spaa{1|2}^3\over
\Spaa{2|4}\Spaa{4|5}...\Spaa{n-1|n}(\Spaa{n|1}+z\Spaa{n|2})}\eean
where $ P_{23}(z)=\ket{2} \left(\bket{2}-z\bket{1}+\bket{3}
{\Spaa{3|\mu}/\Spaa{2|\mu}} \right)$ has been used. Comparing both we arrive
\bean {1\over \Spaa{1|2}\Spaa{3|4}}\left( {\Spaa{n|3}\Spaa{1|2}\over
\Spaa{n|2}} \right)^4 \prod_{l=1}^{v+1} { w_l-z\over w_l- z_\a}=\left(
{\Spaa{3|\mu}\over \Spaa{2|\mu}}\right)^3 {\Spaa{1|2}^3\over
\Spaa{2|4}} \eean
from which one has
\bea \prod_{l=1}^{v+1} { w_l-z\over w_l-
z_\a}=1~~~\label{MHV-P23-1}\eea
here we have used  ${\Spaa{n|3}/\Spaa{n|2}}={\Spaa{3|\mu}/ \Spaa{2|\mu}}={\Spaa{3|4}/ \Spaa{2|4}}$.
(\ref{MHV-P23-1}) holds {\sl when and only when $w_l\to \infty$ under the limit}.
It is tempting to conjecture $w_l\sim {\Spaa{3|1}/\Spaa{3|2}}$.
But we shall see,
roots need not be a rational function and a general prediction cannot be made.

These conclusions can be reached by general analysis as well.
$M_{n-1}( 1^-, P_{23}^{-},4,...,n)$ in (\ref{Pik-Comp-z-2}) has no boundary contributions
(with deformation $\Spba{P_{23}|1}$
and $P_{23}(z)=\ket{2} \left(\bket{2}-z\bket{1}+\bket{3} {\Spaa{3|\mu}/\Spaa{2|\mu}} \right)$
and $p_1(z)=(\ket{1}+z\ket{2})\bket{1}$),
thus we have $\W v+1=0$.
 We reach (\ref{MHV-P23-1}) immediately from (\ref{Pik-check}) as it has only one-cut and $h_2+h_3-h_{23}-1=0$.

\subsubsection{The $P_{12}$ pole}

The limit $\Spbb{2|1}\to 0$ is trivial and we consider only $\Spaa{2|1}\to 0$.
Using $\ket{P_{12}}=\ket{2}\left( \bket{2}+\bket{1}{\Spaa{\mu|1}/\Spaa{\mu|2}}\right)$ and
comparing limits from (\ref{Mz-new}) and direct
factorization we arrive
\bea {1\over \Spaa{2|3}}\left( {\Spaa{n|3}\Spaa{1|2}\over
\Spaa{n|2}} \right)^4 {1\over \Spaa{n|1}+z\Spaa{n|2}}\prod_{l=1}^{v+1} {
w_l-z\over w_l- z_\a}=\left( {\Spaa{\mu|1}\over
\Spaa{\mu|2}}+z\right)^3 {\Spaa{2|3}^3\over \Spaa{n|2}}, \label{5.4} \eea
thus $w_l=-{\Spaa{t_l|1}/\Spaa{t_l|2}}$, $w_l-z_{n1}=
{\Spaa{1|2}\Spaa{t_l|n}/\Spaa{2|n}\Spaa{t_l|2}}$.
Putting it back to (\ref{5.4}) we find
\bea \left( {\Spaa{n|3}\Spaa{1|2}\over \Spaa{n|2}} \right)^4
\prod_{l=1}^4 {\Spaa{2|n}\Spaa{t_l|2}\over
\Spaa{1|2}\Spaa{t_l|n}}=\Spaa{2|3}^4,~~~\to~~ \prod_{l=1}^4
{\Spaa{t_l|2}\over \Spaa{t_l|n}}= {\Spaa{3|2}^4\over
\Spaa{3|n}^4}\eea
The solution is  $t_l=p_3$. As expected, it is the right answer.

\subsection{Example II---The Einstein-Maxwell Theory}

The second example is a theory of photons coupled with gravitons.
In addition to three-point graviton amplitudes,
there are two extra three-point amplitudes
\bea M_3( 1^{-}_\gamma, 2^+_\gamma, 3_g^{-2})= \kappa
{\Spaa{3|1}^4\over \Spaa{1|2}^2},~~~~~ M_3( 1^{-}_\gamma,
2^+_\gamma,3_g^{+2})= \kappa {\Spbb{3|2}^4\over
\Spbb{1|2}^2}~.~~~\label{PG-3-point} \eea
We will take the $\Spba{1|2}$-deformation
\bea \W\la_1\to \W\la_1-z\W\la_2,~~~~~\la_2\to
\la_2+z\la_1~.~~~\label{PG-def} \eea
%

\subsubsection{The four-point amplitude $M_4(1^-_\gamma, 2^+_\gamma, 3_g^{-2},
4_g^{+2})$} 

There are two poles $s_{13}=0$ and $s_{14}=0$ in the recursion relation with boundaries,
but the pole at $s_{14}=0$ gives no contribution under the deformation in (\ref{PG-def}).
Thus we have only one term
\bea M_4(1^-_\gamma, 2^+_\gamma, 3_g^{-2}, 4_g^{+2})(z) =
{\Spbb{2|4}^4\Spaa{1|3}^2\Spaa{2|3}^2\over s_{13} s_{23}^2 }
{\Spbb{3|1}\over \Spbb{3|1}-z\Spbb{3|2}}\prod_{l=1}^{v+1}
{w_l-z\over w_l-z_\a}~~~~\label{PG-middle-1-z} \eea
where $z_\a={\Spbb{3|1}/\Spbb{3|2}}$, $\bket{\WH 1}=
{\bket{3}\Spbb{1|2}/\Spbb{3|2}}$.

\subsubsection*{Poles with $z$-dependence:}

Two poles $P^2_{13}(z)=0$ and $P^2_{14}(z)=0$ exist here.
For the pole $P^2_{13}(z)=0$,  $\Spbb{1(z)|3}$ cannot vanish for arbitrary $z$, but $\Spaa{1|3}$ can.
In the $1^{-1},3^{-2}$ helicity configuration, the factorization limit is trivial
and this pole gives no information on roots at all.
For the pole $P^2_{14}(z)=0$, the $\Spaa{1|4}\to 0$ limit is not trivial and will be discussed carefully.

There are several nontrivial facts for the limit $\Spaa{1|4}\to 0$.
Due to momentum conservation,
$z_\a={\Spbb{3|1}/\Spbb{3|2}}={-\Spaa{2|4}/\Spaa{1|4}}\to \infty$ under the limit.
As the pole $\Spaa{i|k}=0$ discussed in the previous section,
there are potentially extra contributions in $T_{jk}(z)$.
Here, the whole contribution comes totally from this extra possibility.
Poles $s_{14}\to 0$ and $s_{23}\to 0$ occurs simultaneously, as they go to zero at the same time.
This nontrivial kinematics happens only in four-point amplitudes.
Due to this special kinematics, there is in fact no free parameter $z$, as to be shown shortly.

The factorization limit from general consideration is
\bea
I_{direct}& \equiv & \lim_{\Spaa{1|4}\to 0}P_{14}(z)^2 M_4(z)
= M_3^a(1^-(z), 4^{+2}, -P_{14}^{+1}(z)) M_3^h(P_{14}^{-1}(z), 2^+,
3^{-2})\nn
& = & (\Spbb{4|1}-z\Spbb{4|2})^2 \left({\Spaa{\mu|4}\over
\Spaa{\mu|1}}\right)^{-2} (\Spaa{3|2}+z\Spaa{3|1})^2 \left(
{\Spbb{\W \mu|3}\over
\Spbb{\W\mu|2}}\right)^{-2}~~~~\label{PG-middle-2-z} \eea
where we have used
\bea P_{14}(z)=\ket{1}\left(\bket{1}-z\bket{2}+
\bket{4}{\Spaa{\mu|4}\over \Spaa{\mu|1}}
\right),~~~P_{23}(z)=\left(\ket{2}+z\ket{1}+\ket{3} {\Spbb{\W
\mu|3}\over \Spbb{\W\mu|2}}\right) \bket{2} \label{5.10}\eea
Since $\Spaa{1|4}\to 0$ implies $P_{23}(z)^2\to 0$,
one must then have either $\ket{2(z)}\sim \ket{3}$ or $\bket{2}\sim \bket{3}$ for all $z$
and the sensible choice is $\bket{2}\sim \bket{3}$.

To get the other factorization limit, we rewrite (\ref{PG-middle-1-z}) as
\bean M_4(1^-_\gamma, 2^+_\gamma, 3_g^{-2}, 4_g^{+2})(z) & = &
{\Spbb{2|4}^2\Spaa{1|3}^4\over s_{13} \Spaa{1|4}^2 }
{\Spbb{3|1}\over \Spbb{3|1}-z\Spbb{3|2}}\prod_{l=1}^{v+1} {w_l-z\over
w_l-z_\a}\eean
As $z_\a\to \infty$, $w_l-z_\a \to -z_\a$.
To get a finite factorization limit, we need {\sl one and only one root}.
Putting all together we obtain
\bea I_{BCFW}&= & \lim_{\Spaa{1|4}\to 0}P_{14}(z)^2 M_4(z)=
{\Spbb{2|4}^2\Spaa{1|3}^4\over s_{13}  }
{\Spbb{3|1}(\Spbb{4|1}-z\Spbb{4|2})\over \Spbb{3|1}} {(w-z)\over
\Spaa{2|4}} \eea
where $\Spbb{2|3}\to 0$ is used.

Superficially, one fails to see that $I_{direct}=I_{BCFW}$ since one is a polynomial of $z$ of degree 4
while the other of degree 2.
They are indeed the same, due to the special kinematics of four particles, as we see presently.
From $P_{14}(z)=-P_{23}(z)$, one has
\bea \ket{1} & = & \a \left(\ket{2}+z\ket{1}+\ket{3} {\Spbb{\W
\mu|3}\over
\Spbb{\W\mu|2}}\right),~~~~\bket{2}=-\a^{-1}\left(\bket{1}-z\bket{2}+
\bket{4}{\Spaa{\mu|4}\over \Spaa{\mu|1}} \right)\eea
which is true if and only if
\bea {\Spbb{\W \mu|3}\over \Spbb{\W\mu|2}}= -{\Spaa{1|2}\over
\Spaa{1|3}},~~~~{\Spaa{\mu|4}\over \Spaa{\mu|1}}=-{\Spbb{2|1}\over
\Spbb{2|4}}\eea
and
\bea \a^{-1}= z+ {\Spaa{3|2}\over \Spaa{3|1}}
=\left(z-{\Spbb{4|1}\over \Spbb{4|2}}
\right)^{-1}~~~\label{a-cond}\eea
The condition (\ref{a-cond}) is in fact very tricky.
${\Spaa{3|2}/\Spaa{3|1}}=-{\Spbb{4|1}/\Spbb{4|2}}$ can be obtained from momentum conservation,
thus (\ref{a-cond}) gives a relation between $z$ and external momenta.
In other words, due to momentum conservation, $z$ is not a variable under the factorization limit.
(\ref{a-cond}) can be solved by
\bea z+ {\Spaa{3|2}\over \Spaa{3|1}}= \kappa= z-{\Spbb{4|1}\over
\Spbb{4|2}},~~~~\kappa=\pm 1~.\eea
Using it we can simplify
\bea I_{BCFW}& = &   -{\Spaa{2|4}^4\Spaa{1|3}^4\over s_{13}^2}
\kappa (w-z),~~~I_{direct}  =  {\Spaa{2|4}^4\Spaa{1|3}^4\over
s_{12}^2}\eea
Finally because  $s_{12}=-s_{13}$, we have
\bea  w-z= -\kappa, \to w= {\Spbb{4|1}\over \Spbb{4|2}},~~~ {w\over
w-z_{13}}=-{s_{23}\over s_{12}} \label{5.17}\eea
Putting it back to (\ref{PG-middle-1-z}) we will get the right
amplitude.

\subsubsection*{Poles without $z$-dependence}

Now there is only one pole $s_{12}=0$.
From the direct factorization limit, one has
\bea & & \lim_{s_{12}\to 0} s_{12} M_4(z) \nn
 & = & M_3^a(1^-,
2^+, -P_{12}^{+2}) M_3^h(P_{12}^{-2}, 3^{-2}, 4^{+2})+M_3^h(1^-,
2^+, -P_{12}^{-2}) M_3^a(P_{12}^{+2}, 3^{-2}, 4^{+2})\nn
& = &   {\Spbb{1|2}^2\Spaa{1|3}^6\over \Spaa{3|4}^2\Spaa{4|1}^2} +
{\Spbb{2|4}^6 \Spaa{1|2}^2\over
\Spbb{3|4}^2\Spbb{3|2}^2}~~~\label{M41-P12-1-z}\eea
where the first term is from the limit $\Spaa{1|2}\to 0$
while the second from the limit $\Spbb{1|2}\rightarrow 0$.
(\ref{M41-P12-1-z}) {\sl does not depend on $z$ at all}.

Now identify (\ref{M41-P12-1-z}) with (\ref{PG-middle-1-z}) after multiplying the latter by $s_{12}$.
The $z$-independence of (\ref{M41-P12-1-z}) means that the factor ${\Spbb{3|1}/\Spbb{3|2}}-z$ in
denominator of (\ref{PG-middle-1-z}) should be canceled by one factor $w-z$.
That is, there is one root.
We may work with two different limits,
namely  $\ket{1}\sim \ket{2}, \bket{3}\sim\bket{4}$ or $\ket{3}\sim \ket{4}, \bket{1}\sim \bket{2}$.
The natural  choice is $w={\Spbb{4|1}/\Spbb{4|2}}$, the same result as given in (\ref{5.17}).
To work out the matching factors, careful kinematic analysis should be
carried out as for the $\Spaa{1|4}\to 0$ limit.

\subsubsection{The four-point amplitude $M_4(1^-_\gamma, 2^+_\gamma, 3^-_\gamma, 4^+_\gamma)$}

There is only one pole $s_{14}$ in the recursion relation with boundaries (cut $s_{13}$ yields a null contribution here).
With the deformation in (\ref{PG-def}),  the boundary amplitude can be written as
\bea M_4(1^-_\gamma, 2^+_\gamma, 3^-_\gamma, 4^+_\gamma)(z) =
{\Spbb{2|4}^2\Spaa{1|3}^2\over  \Spaa{1|4}(\Spbb{4|1}-z\Spbb{4|2})}
\prod_{l=1}^{v+1} {w_l-z\over w_l-z_\a}~~~~\label{PG42-middle-1-z}
\eea
where $z_\a={\Spbb{4|1}/\Spbb{4|2}}$, $\bket{\WH 1}= \bket{4}{\Spbb{1|2}/\Spbb{4|2}}$.

\subsubsection*{Poles with $z$-dependence:}

Here one has two poles at $P^2_{13}(z)=0$ and $P^2_{14}(z)=0$.
For the pole $P^2_{13}(z)=0$,  $\Spbb{1(z)|3}\to 0$ can not be true for arbitrary $z$, but $\Spaa{1|3}\to 0$ can.
In the $1^-,3^-$ helicity configuration, the factorization limit is trivial
and this pole gives no information on roots at all.
For the pole $P^2_{14}(z)$, the $\Spaa{1|4}\to 0$ limit is not trivial.
Similar to discussions in the previous subsection,
due to momentum conservation and the special kinematics of four particles,
$z$ cannot vary, as to be discussed presently.

The factorization limit from (\ref{PG42-middle-1-z}) is
\bea
I_{BCFW}&= & \lim_{\Spaa{1|4}\to 0}P_{14}(z)^2 M_4(z)=\Spbb{2|4}^2\Spaa{1|3}^2\prod_{l=1}^{v+1} {w_l-z\over w_l-z_\a}
 \eea
and the factorization limit from general consideration is
\bea
I_{direct}& \equiv & \lim_{\Spaa{1|4}\to 0}P_{14}(z)^2 M_4(z)\nn
&=& M_3^a(1^-(z), 4^+, -P_{14}^{+2}(z)) M_3^h(P_{14}^{-2}(z), 2^+, 3^-)
   + M_3^a(1^-(z), 4^+, -P_{14}^{-2}(z)) M_3^h(P_{14}^{+2}(z), 2^+, 3^-)\nn
& = & (\Spbb{4|1}-z\Spbb{4|2})^2
(\Spaa{3|2}+z\Spaa{3|1})^2~~~~\label{PG42-middle-2-z} \eea
where $P_{14}(z)$ and $P_{23}(z)$ are listed in (\ref{5.10}).
Following reasonings after (\ref{5.10}), we have to choose $\bket{2}\sim \bket{3}$.

Setting $I_{direct}=I_{BCFW}$ and $w\rightarrow z_\a$,
the dominator in $I_{BCFW}$ must go to zero if the $z$-dependent part is consistent.
The subtlety again resides in the special kinematics.
$I_{direct}$ can actually be written as
\bea
I_{direct}& = & (\Spbb{4|1}-z\Spbb{4|2})^2 (\Spaa{3|2}+z\Spaa{3|1})^2\nn
&=&\Spbb{4|2}^2 \Spaa{1|3}^2
\eea
Compared with the $I_{BCFW}$, one sees that $w\rightarrow \infty$ under this limit.

\subsubsection*{Poles without $z$-dependent}

There is only pole $s_{12}=0$ in this case.
The factorization limit from general theory is
\bea & & \lim_{s_{12}\to 0} s_{12} M_4(z) \nn
 & = & M_3^a(1^-,2^+, -P_{12}^{+2}) M_3^h(P_{12}^{-2}, 3^-, 4^+)
 +M_3^h(1^-,2^+, -P_{12}^{-2}) M_3^a(P_{12}^{+2}, 3^-, 4^+)\nn
& = &   \Spbb{1|2}^2\Spaa{3|4}^2 +
\Spaa{1|2}^2\Spbb{3|4}^2~~~\label{M42-P12-1-z}\eea
where the first term is from the limit $\Spaa{1|2}\to 0$ while the second from the limit $\Spbb{1|2}\rightarrow 0$.
(\ref{M42-P12-1-z}) {\sl does not depend on $z$ at all}.

Now identify (\ref{M42-P12-1-z}) with (\ref{PG42-middle-1-z}) after multiplying the latter by $s_{12}$.
The $z$-independence of (\ref{M42-P12-1-z}) means that the factor ${\Spbb{4|1}/ \Spbb{4|2}}-z$ in
denominator of (\ref{PG42-middle-1-z}) should be canceled by one factor $w-z$ under this limit.
That is, there is only one root.

There are two different limits,
$\ket{1}\sim \ket{2}, \bket{3}\sim\bket{4}$ and $\ket{3}\sim \ket{4}, \bket{1}\sim \bket{2}$.
The natural  choice is $w={\Spbb{3|1}/ \Spbb{3|2}}=-{\Spaa{4|2}/ \Spaa{4|1}}$.
This gives naturally $w\rightarrow \infty$ as $\Spaa{1|4}\rightarrow 0$.

\subsubsection{The five-point amplitude $M(1^{-1}_\gamma, 2^{+1}_\gamma, 3^{-1}_\gamma, 4^{+1}_\gamma, 5^{-2}_g)$}

Written as a BCFW expansion,
the five-point amplitude $M(1^{-1}_\gamma, 2^{+1}_\gamma, 3^{-1}_\gamma, 4^{+1}_\gamma, 5^{-2}_g)$
can be deformed as
\bea
M_5(z) &= & M_3( \WH 1^{-1}, 5^{-2}, \WH P_{15}^{+1}) {1\over s_{15}(z)}
M_4(\WH P_{15}^{-1},\WH 2^+, 3^-, 4^+) \prod_{l=1}^{v+1} {w_l-z\over w_l-z_{15}} \nn
& & +M_3(\WH 1^-, 4^+, P_{14}^{-2}) {1\over s_{14}(z)} M_4(\WH P_{14}^{+2},
\WH 2^+, 3^-, 5^{-2}) \prod_{l=1}^{v+1} {w_l-z\over w_l-z_{14}} \nn
&=& { \Spaa{1|5} \Spaa{3|4} \Spbb{3|5} \Spbb{2|4}^5\over
\Spbb{4|3} \Spbb{2|3} \Spbb{4|5} \Spbb{2|5}^2 (\Spbb{5|1}-z\Spbb{5|2})}\prod_{l=1}^{v+1}{w_l-z\over
w_l-z_{15}}\nn
& & + {\Spaa{1|4}\Spaa{3|5}\Spbb{2|4}^4\Spbb{4|3}\over
(\Spbb{4|1}-z\Spbb{4|2})\Spbb{5|3}\Spbb{2|3}\Spbb{4|5}\Spbb{5|2}}\prod_{l=1}^{v+1}{w_l-z\over
w_l-z_{14}}~~~~\label{PG5-middle-1-z} \eea
In the first term $z_{15}={\Spbb{5|1}/\Spbb{5|2}}$, $\bket{\WH 1}= {\bket{5}\Spbb{1|2}/ \Spbb{5|2}}$
and in the second term $z_{14}={\Spbb{4|1}/\Spbb{4|2}}$,$\bket{\WH 1}= {\bket{4}\Spbb{1|2}/\Spbb{4|2}}$.

\subsubsection*{Poles with $z$-dependence:}

There are four poles resulting from the vanishing of $P_{15}(z)$, $P_{14}(z)$, $P_{25}(z)$ and $P_{23}(z)$, respectively.
Other possible poles give trivial factorization limits.
For poles $P_{15}(z)=0$ and $P_{14}(z)=0$,
both $\Spaa{1|5}\to 0$ and $\Spaa{1|4}\to 0$ make the factorization limit trivial.
They yield nothing at all.
$\Spbb{2|5}\to 0$ and $\Spbb{2|3}\to 0$ are nontrivial limits,
but $\Spaa{2(z)|5}\to 0$ and $\Spaa{2(z)|3}\to 0$ cannot be true for arbitrary z.

Consider $\Spbb{2|5}\to 0$ first.
The factorization limit from general theory is
\bea
I_{direct}& \equiv & \lim_{\Spbb{2|5}\to 0}P_{25}(z)^2 M_5(z)\nn
&=& M_4(1^-(z), P_{25}^+(z),3^-,4^+ ) M_3(P_{25}^-(z), 2^+(z), 5^{-2})\nn
& =&{\Spbb{\mu|2}\over \Spbb{\mu|5}}^2
{\Spaa{1|3}^3\Spbb{4|2}^2(\Spaa{5|2}+z\Spaa{5|1})^2(\Spbb{3|1}-z\Spbb{3|2})
\over
\Spaa{1|4}\Spbb{1|2}(\Spaa{1|2}+\Spaa{1|5}\Spbb{\mu|5}/\Spbb{\mu|2})(\Spbb{4|1}-z\Spbb{4|2})}
~~~~\label{PG5-middle-2-z} \eea where we have used
\bea
P_{25}(z)=\bket{2}\left(\ket{2}+z\ket{1}+\ket{5}{\Spbb{\mu|5}\over \Spbb{\mu|1}}\right)
\eea
The factorization limit from (\ref{PG5-middle-1-z}) is
\bea
I_{BCFW}&= & \lim_{\Spbb{2|5}\to 0}P_{25}(z)^2 M_4(z)\nn
&=&\left\{{ \Spaa{1|5}\Spaa{3|4} \Spbb{3|5} \Spbb{2|4}(\Spaa{5|2}+z\Spaa{5|1})\over
\Spbb{4|3}  \Spbb{2|5}(\Spbb{5|1}-z\Spbb{5|2})}
\prod_{l=1}^{v+1}{w_l-z\over w_l-z_{15}}\right.\nn
& &\left. + {\Spaa{1|4}\Spaa{3|5}\Spbb{3|4}(\Spaa{5|2}+z\Spaa{5|1})\over
\Spbb{5|3}(\Spbb{4|1}-z\Spbb{4|2})}
\prod_{l=1}^{v+1}{w_l-z\over w_l-z_{14}}\right\}{\Spbb{2|4}^4\over\Spbb{2|3} \Spbb{4|5} }
 \eea
Identifying $I_{BCFW}$ with $I_{direct}$ and noticing that $z_{15}\to \infty$,
one gets one root in this limit:
$w=-{\Spbb{3|1}/ \Spbb{3|2}}$.

The analysis of $\Spbb{2|3}\to 0$ is analogous that of $\Spbb{2|5}\to 0$.
To make the degrees of $z$ identical in these two factorization limits,
the root $w$ here has to become infinity under $\Spbb{2|3}\to 0$.
This implies that there may be an factor $\Spbb{2|3}$ in the denominator.

\subsubsection*{Poles without $z$-dependent}

Now the $z$-independent limits.
Poles without $z$-dependence include two categories: $P^2_\a\to 0 $($\a \subset I$ or $\a \subset \O{1,2,I}$ )  and $P^2_{12}\to 0$.
And there are three kinds of $P^2_\a\to 0 $ limits:
(1) from the collinear limit $s_{34}$ (where $\W \la_3\sim \W\la_4$) one obtains a single root
$w=-{\Spaa{2|5}/\Spaa{1|5}}={\Spbb{1|P_{34}}/\Spbb{2|P_{34}}}$ with $M_4(1^-,2^+,P_{34}^{+2}, 5^{-2})$ in direct factorization limit;
(2) from the collinear limit $s_{35}$ (where $\W\la_3\sim \W \la_5$) one gets a single root
$w=-{\Spaa{2|4}/\Spaa{1|4}}={\Spbb{1|P_{35}}/\Spbb{2|P_{35}}}$ with $M_4(1^-,2^+, P_{35}^{-1}, 4^{+})$ in direct limit part;
(3) from the collinear limit $s_{45}$ (where $\W\la_4\sim \W\la_5$) one finds a single root
$w={\Spbb{1|3}/\Spbb{2|3}}$with  $M_4(1^-,2^+, 3^{-1},P_{45}^{+})$ in direct limit part.
From these, we deduce the root's expression as
\bea
w={\Spbb{1|3}\over \Spbb{2|3}}(1+ f_1
\Spbb{3|4}\Spbb{3|5}\Spbb{4|5})
\eea
with following requirements:
(1) $f_1 \Spbb{3|4}\Spbb{3|5}\Spbb{4|5}$ should be dimensionless;
(2) $f_1 \Spbb{3|4}\Spbb{3|5}\Spbb{4|5}$ should be helicity neutral for all external particles;
(3) the ${w/ (w-z_\a)}$ should not produce un-physical pole when we collect all results.
One is then led to the natural choice $w={\Spbb{1|3}/\Spbb{2|3}}$.

Now the roots under the limit $P^2_{12}\to 0$.
$\Spaa{1|2}\to 0$ will lead to a trivial result, while $\Spbb{1|2}\to 0$ could result in a solution.
Moreover, $z_{15}\sim z_{14}\to 1$ under this limit.
Comparing two different factorization limit under the limit $\Spbb{1|2}\to 0$,
one finds that $w=1$.

Together with above discussions, the root in $M(z)$ is $w={\Spbb{3|1}/\Spbb{3|2}}$.
Plugging it back to $M(z=0)$, one obtains the same amplitude as in \cite{Benincasa:2011kn}.

\subsection{Example III--- The six-gluon amplitude $M_6(1^-,2^-,3^-,4^+,5^+,6^+)$}

In previous examples we can solve roots with the help of their
factorization limits.
One naturally asks whether this is possible in general.
In this subsection, we will use the example of six-gluon amplitude
$M_6(1^-,2^-,3^-,4^+,5^+,6^+)$ to show that, generally, knowing
values  of roots under all factorization limits is not enough
 to solve them. In fact, values of roots are not even simple
 rational functions of spinor contraction $\Spaa{~|~}$,
 $\Spbb{~|~}$.

The six-gluon amplitude $M_6(1^-,2^-,3^-,4^+,5^+,6^+)$ is given by
\bea M_6(1^-,2^-,3^-,4^+,5^+,6^+)={1\over
\Spab{5|3+4|2}}\bigg({\Spab{1|2+3|4}^3\over
[23][34]\Spaa{5|6}\Spaa{6|1}(p_2+p_3+p_4)^2}\nn
+{\Spab{3|4+5|6}^3\over
[61][12]\Spaa{3|4}\Spaa{4|5}(p_3+p_4+p_5)^2}\bigg),~~~\label{A6-split-hel}
\eea
For our purpose we will use the
deformation-$\Spba{5|3}$\footnote{For deformation $\Spba{4|3}$,
there is no pole and the recursion relation should be modified accordingly.}.
\bea \ket{3} \to \ket{3}+z
\ket{5},~~\bket{5}\to\bket{5}-z\bket{3}.\eea
Under this deformation, the boundary BCFW recursion relation gives
following $z$-dependent amplitudes
 \bea
& &
M_6(1^-,2^-,\wh{3}(z)^-,4^+,\wh{5}(z)^+,6^+)\nn&=&{\Spba{6|5+3|4}^3
\Spaa{3|5}^3 \over
\Spba{2|3+4|5}\Spaa{4|5}^4\Spbb{6|1}\Spbb{1|2}P_{345}^2(\Spaa{3|4}+z\Spaa{5|4})
}\prod_l{w_l-z\over w_l-z_{34}}\nn
&+&{\Spba{4|2+3|5}^3\Spba{3|5+6|1}^3\over
\Spba{2|3+4|5}\Spba{3|2+4|5}^3\Spbb{2|3}\Spbb{4|3}\Spaa{5|6}\Spaa{6|1}(P_{234}^2+z\Spba{3|2+4|5})}\prod_l{w_l-z\over
w_l-z_{234}}.\nn\label{6nmhv-z}\eea
where $z_{34}=-{\Spaa{4|3}/\Spaa{4|5}}$ and
$z_{234}=-{P_{234}^2/\Spba{3|2+4|5}}$.

The pole structure of six-gluon amplitude is the following. There are
three three-particles poles, $s_{123}=s_{456}$, $s_{234}=s_{561}$, $s_{345}=s_{612}$.
Among them, the split helicity configuration, $s_{123}=s_{456}$ is trivial. For two
particle poles we need to consider the holomorphic and anti-holomorphic part.
After splitting helicity configurations, nontrivial channels are
\bea
\Spbb{1|2},~~\Spbb{2|3},~~~\Spaa{3|4},~~~\Spbb{3|4},~~\Spaa{4|5},~~\Spaa{5|6},
~~~\Spaa{6|1},~~~\Spbb{6|1}\eea
%

\subsubsection{Poles under factorization limits without $z$-dependence}

Here one has one three-particle channel $P_{612}$
and three two-particle channels $\Spbb{1|2}$, $\Spaa{6|1}$, $\Spbb{6|1}$.
Since $3,5$ are not nearby, we do not have the pole $P_{35}$, as mentioned before.

{\bf Pole $P_{612}$:} When $P_{216}^2\to 0$, factorization limit leads to
\bean & & M_4(6^+,1^-,
2^-,-P_{612}^+)M_4(P_{612}^-,\wh{3}^-,4^+,\wh{5}^+)={\Spaa{1|2}^3\Spbb{4|5-z3}^3\over
\Spbb{3|4}\Spaa{6|1}\Spab{2|P_{612}|3}\Spba{5-z3|P_{612}|6}}\eean
which leads to triple roots $ w_l^{(3)}= {\Spbb{4|5}/\Spbb{4|3}}$.
This shows that we could find roots without working out detailed comparison,
evidencing certain power of $z$-dependent factorization limits.

{\bf Pole $\Spbb{1|2}$:} The factorization limit is
\bean  M_3(1^-,2^-, -P_{12}^+)
M_5(P_{12}^-,\wh{3}^-,4^+,\wh{5}^+,6^+)=
\Spaa{1|2}{\Spbb{\mu|1}\over \Spbb{\mu|2}} {(-)
(\Spba{\mu|1+2|3}+z\Spba{\mu|1+2|5})^3\over \Spaa{3+z5|4}
\Spaa{4|5}\Spaa{5|6} \Spba{\mu|1+2|6}\Spbb{\mu|1}^2}\eean
which leads to triple roots $
w_l^{(3)}=-{\Spba{\mu|1+2|3}/\Spba{\mu|1+2|5}}$.

{\bf Pole $\Spaa{6|1}$:} The factorization limit gives
\bean  M_3(6^+,1^-,
-P_{16}^+)M_5(P_{16}^-,2^-,\wh{3}^-,4^+,\wh{5}^+) =
{\Spbb{1|6}\Spaa{\mu|1}^3 \Spbb{4|5-z 3}^3\over
\Spaa{\mu|6}\Spba{5-z3|1+6|\mu}\Spba{2|1+6|\mu}\Spbb{2|3}\Spbb{3|4}}\eean
which leads to triple roots $w_l^{(3)}={\Spbb{4|5}/
\Spbb{4|3}}$.

{\bf Pole $\Spbb{6|1}$:} The factorization limit gives
\bean  M_3(6^+,1^-,
-P_{16}^-)M_5(P_{16}^+,2^-,\wh{3}^-,4^+,\wh{5}^+)= {\Spbb{\mu|6}^3
\Spaa{1|6}\Spaa{2|3+z5}^3\over
\Spbb{\mu|1}\Spab{5|1+6|\mu}\Spab{2|1+6|\mu} \Spaa{3+z 5|4}}\eean
which leads to triple roots  $
w_l^{(3)}=-{\Spaa{2|3}/ \Spaa{2|5}}$.

\subsubsection{Poles under factorization limits with $z$-dependence}

For this type, pole $P_{234}$ does not have $z$-dependent
factorization limit  and need not to be discussed.
We are left
with five two particle poles $\Spbb{2|3}$, $\Spaa{3|4}$,
$\Spbb{3|4}$, $\Spaa{4|5}$, and $\Spaa{5|6}$. Among them,
$\Spbb{3|4}$ is automatically satisfied by recursion relation.

{\bf Pole $\Spbb{2|3}$:} The direct factorization gives
\bean
M_3(2^-,\wh{3}^-,-\wh{P}_{23}^+)M_5(\wh{P}_{23}^-,4^+,\wh{5}^+,6^+,1^-)=
{(\Spaa{2|3}+z\Spaa{2|5})(\Spba{\mu|2+3|1}+z\Spaa{1|5}\Spbb{3|\mu})^3
\over\Spbb{2|\mu}\Spbb{\mu|3}\Spaa{4|5}\Spaa{5|6}\Spaa{6|1}
(\Spba{\mu|2+3|4}+z\Spaa{5|4}\Spbb{\mu|3})}.\eean
Compared with contribution from the second term of (\ref{6nmhv-z}),
we find triple roots
$w_l^{(3)}=-{\Spba{\mu|2+3|1}/\Spbb{\mu|3}\Spaa{5|1}}$.

{\bf Pole  $\Spbb{3|4}$:} The direct factorization gives
\bean M_3(\wh{3}^-,4^+,-\wh{P}_{34}^-)M_5
(\wh{P}_{34}^+,\wh{5}^+,6^+,1^-,2^-) =
{\Spbb{\mu|4}^3\Spaa{1|2}^3(\Spaa{3|4}+z\Spaa{5|4})\over
\Spbb{\mu|3}\Spba{\mu|3+4|5}\Spaa{5|6}\Spaa{6|1}(-\Spba{\mu|4+3|2}+z\Spaa{2|5}\Spbb{\mu|3})},\eean
which does not have nontrivial $z$-dependence (factor $(\Spaa{3|4}+z\Spaa{5|4})$ comes from $s_{\WH 34}$).
From our previous discussions, it can happen when and only when $w\to \infty$ under the limit.

{\bf Pole $\Spaa{4|5}$:} The factorization limit gives
\bean
M_5(6^+,1^-,2^-,\wh{3}^-,\wh{P}_{45}^+)M_3(-\wh{P}_{45}^-,4^+,\wh{5}^+)=
{(\Spba{6|4+5|\mu}-z\Spbb{3|6}\Spaa{\mu|5})^3(\Spbb{5|4}-z\Spbb{3|4})\over
\Spaa{\mu|4}\Spaa{\mu|5}\Spbb{6|1}\Spbb{1|2}\Spbb{2|3}\Spba{3|4+5|\mu}},\eean
which leads to triple roots  $w_l^{(3)}=-{\Spba{6|4+5|\mu}/\Spbb{6|3}\Spaa{\mu|5}}$.

{\bf Pole $\Spaa{5|6}$:} The factorization limit gives
\bean M_5(1^-,2^-,\wh{3}^-,4^+,
\wh{P}_{56}^+)M_3(-\wh{P}_{56}^-,\wh{5}^+,6^+)=
{(\Spba{4|5+6|\mu}+z\Spbb{4|3}\Spaa{\mu|5})^3(\Spbb{6|5}-z\Spbb{3|5})\over
\Spaa{\mu|6}\Spaa{\mu|5}\Spbb{1|2}\Spbb{2|3}\Spbb{3|4}(\Spba{1|5+6|\mu}-z\Spbb{3|1}\Spaa{\mu|5})},\eean
which leads to triple roots  $w_l^{(3)}=-{\Spba{4|5+6|\mu}/ \Spbb{4|3}\Spaa{\mu|5}}$.

\subsubsection{True values of roots}

So far, roots have been found under various factorization limits.
We wish to find roots without taking the limits, to reproduce known results in (\ref{A6-split-hel}).
However, without using the known result (\ref{A6-split-hel}), we are not able to do so.
To show why it is so difficult to solve roots with the help of factorization limits,
we now discuss roots directly from (\ref{A6-split-hel}).

The numerator from expression (\ref{A6-split-hel}) is given by
\bea N & = & T_1+ T_2 \nn
T_1 & = & -\Spaa{4|5}\Spbb{2|1}\Spbb{6|1}
s_{345}\Spaa{4|5}\Spaa{1|5}^3\Spbb{4|3}^3\left(z+{\Spaa{3|4}\over\Spaa{5|4}}\right)
\left(-{\Spba{4|P_{23}|1}\over \Spaa{1|5}\Spbb{4|3}}+z\right)^3\nn
T_2 & = &
-\Spaa{1|6}\Spaa{5|6}\Spbb{3|2}\Spbb{4|3}\Spab{5|P_{234}|3}\Spab{5|P_{345}|6}^3\left(
{ s_{234}\over \Spab{5|P_{234}|3}}+z \right)
\left({\Spab{3|P_{345}|6}\over \Spab{5|P_{345}|6}}+z
\right)^3~~~\label{6-num}\eea
From (\ref{6-num}) we can read out values of roots under
various factorization limits
\bea   &&\Spbb{1|2}\to
0,~~w_l^{(3)}=-{\Spba{\mu|1+2|3}\over\Spba{\mu|1+2|5}}=-{\Spba{6|4+5|3}\over
\Spba{6|3+4|5}},\nn
&&\Spaa{1|6}\to0,~~ w_l^{(3)}={\Spbb{4|5}\over
\Spbb{4|3}}={\Spba{4|2+3|1}\over \Spbb{4|3} \Spaa{1|5}},\nn
&&\Spbb{1|6}\to0,~~ w_l^{(3)}=-{\Spaa{2|3}\over
\Spaa{2|5}}=-{\Spba{6|4+5|3}\over \Spba{6|3+4|5}},\nn
&&P_{216}^2\to 0,~~w_l^{(3)}={\Spbb{4|5}\over
\Spbb{4|3}}=-{\Spba{6|4+5|3}\over \Spba{6|3+4|5}},\nn
&&\Spbb{2|3}\to
0,~~w_l^{(3)}=-{\Spba{\mu|2+3|1}\over\Spbb{\mu|3}\Spaa{5|1}}={\Spba{4|2+3|1}\over
\Spbb{4|3} \Spaa{1|5}},\nn
&&\Spbb{3|4}\to 0,~~w_l^{(3)}\to \infty,\nn
&&\Spaa{5|6}\to 0,~~w_l^{(3)}=-{\Spba{4|5+6|\mu}\over
\Spbb{4|3}\Spaa{\mu|5}}={\Spba{4|2+3|1}\over \Spbb{4|3}
\Spaa{1|5}},\nn
&&\Spaa{5|4}\to 0,~~w_l^{(3)}=-{\Spba{6|4+5|\mu}\over
\Spbb{6|3}\Spaa{\mu|5}}=-{\Spba{6|4+5|3}\over
\Spba{6|3+4|5}}.~~\label{6-limit-zero}\eea

The reason why we obtained simple rational expressions for roots is that one of $T_1, T_2$ will be zero under these factorization limits.
However, for general momentum configurations,
$T_1$ and $T_2$ are not zero, thus we have to solve roots of degree four polynomial.
The analytic expression for roots is very complicated and it is not
{\sl rational function of spinor}
\footnote{We have checked this using numerical method by setting all spinor components to be integer number.}.
Because the irrationality, even with information given in (\ref{6-limit-zero}), it is very hard to find explicit expressions.

\section{Conclusion}

Understanding nontrivial boundary contributions is important in the application of BCFW recursion relations.
In \cite{Benincasa:2011kn}, they were translated to discussion of roots of amplitudes.
In this paper, we have investigated some aspects of roots.

First we re-derived BCFW recursion relations with boundary contributions from a different perspective.
Then we  generalized the factorization limits  to $z$-dependent ones,
where the behavior of roots under the limit can be seen more clearly.
The merits or the demerits of these analyses was illustrated by examples.
One sees that information extracted from roots under various factorization limits is valuable,
but not powerful enough to guarantee explicit expressions of roots.
Our analysis has not been conclusive.
We have the feeling that it may not be practical to find the boundary contributions through roots,
though it does help to clarify certain theoretical issues, as shown in this paper and in \cite{Benincasa:2011pg}.

Roots of amplitudes have not been discussed extensively in quantum field theories.
Their roles are still obscure.
It may help to understand quantum field theories if they can get more thorough scrutinization.
And we believe that they deserve the attention.

\subsection*{Acknowledgements}

 We are supported by fund from
Qiu-Shi, the Fundamental Research Funds for the Central Universities
with contract number 2010QNA3015, National Basic Research Program of China (2010CB833000),
as well as Chinese NSF funding under contract Nos.10875104, 11031005, 10875103, 11135006, 11125523.


\end{document}